\documentclass[letterpaper,twocolumn,10pt]{article}
\pdfoutput=1
\def\isReadyToSubmit{0}   % mark 1 before submission (draft or camera ready).
\def\isCameraReady{0}     % mark 1 for camera ready.
\def\isAnonymousSubmission{0}  % mark 1 if this is an anonymous submission.
\def\isTR{1}  % mark 1 if this is the techreport version.
\def\sandy{0}             % mark 1 if for Sandy.

\usepackage[all=normal, title, sections, paragraphs, indent, lists, floats]{savetrees}

\usepackage{graphicx}
\usepackage{subfigure}
\usepackage{amsmath,amssymb}
\usepackage{color}
\usepackage[hyphens]{url}
\usepackage{listings}
\usepackage[normalem]{ulem}
\usepackage{floatrow}
\usepackage{wrapfig}
\usepackage{xspace}
\usepackage{extarrows}
\usepackage{sidecap}

\ifnum\sandy=1
  \usepackage{setspace}
\fi

\ifnum\isAnonymousSubmission=1
  \ifnum\isCameraReady=1
    \def\includeAuthor{1}
  \else
    \def\includeAuthor{0}
  \fi
\else
  \def\includeAuthor{1}
\fi

\newcommand{\ie}{i.e.,~}
\newcommand{\eg}{e.g.,~}
\def\F{Figure~}

\newcommand{\ar}[3]{} %% something bibtex is doing - ignore it
\include{latexdefs}

\newcommand{\headingg}[1]{\noindent{\bf{#1}}} % at beginning of section -- no extra space
\newcommand{\heading}[1]{{\vspace{2pt}\noindent\bf{#1}}} % inside section
\newcommand{\headingi}[1]{{\vspace{0pt}\em{#1}}} % inside section
{\makeatletter
 \gdef\xxxmark{%
   \expandafter\ifx\csname @mpargs\endcsname\relax % in minipage?
     \expandafter\ifx\csname @captype\endcsname\relax % in figure/caption?
       \marginpar{\textcolor{red}{xxx~}}% not in a caption or minipage, can use marginpar
     \else
       \textcolor{red}{xxx~}% notice trailing space
     \fi
   \else
     \textcolor{red}{xxx~}% notice trailing space
   \fi}
 \gdef\xxx{\@ifnextchar[\xxx@lab\xxx@nolab}
 \long\gdef\xxx@lab[#1]#2{{\bf [\xxxmark \textcolor{red}{#2} ---{\sc #1}]}}
 \long\gdef\xxx@nolab#1{{\bf [\xxxmark \textcolor{red}{#1}]}}
 \ifnum\isReadyToSubmit=1
   \long\gdef\xxx@lab[#1]#2{}\long\gdef\xxx@nolab#1{}%
 \fi
}

\newcommand{\escapexxx}[1]{
  \ifnum\isReadyToSubmit=0
    \textbf{[\textcolor{red}{xxx---#1}]}
  \fi
}

\definecolor{grey}{rgb}{0.5,0.5,0.5}

\newcommand{\code}[1]{\tt{#1}}

\newcommand{\midwor}[1]{\;\textnormal{ #1 }\;} % text in math mode
\newcommand{\vf}{\,,\,} % a comma with space
\newcommand{\ff}{\,.} % a dot 
\newcommand{\lset}{\left\{\left.\;}

\newcommand{\dimset}{\right.\;\left|\;}
\newcommand{\rset}{\;\right.\right\}}

\newcommand{\NatInt}{\mathbb{N}} % natural integer numbers
\newcommand{\probaof}[1]{\mathbb{P}\left.\left[#1\right.\right]} % Probability of an event
\newcommand{\cond}{\right|\left.} % Condition on an event 
\newcommand{\expec}[1]{\mathbb{E}\left[#1\right]} % Expectation 
\newenvironment{disarray}%
 {\everymath{\displaystyle\everymath{}}\array}%
 {\endarray}

\newcommand{\punt}{p_{\emptyset}}
\newcommand{\pin}{p_{\textrm{in}}}
\newcommand{\pout}{p_{\textrm{out}}}
\newcommand{\aacct}{A_k}

\newtheorem{thm}{Theorem}

\newtheorem{lem}{Lemma}
\newenvironment{proof}{\textbf{Proof:}}{$\square$}

\usepackage{times}

\usepackage{usenix}

\usepackage{hyperref}
\usepackage{breakurl}
\usepackage[hyphens]{url}
\def\UrlBreaks{\do\/\do-}
\expandafter\def\expandafter\UrlBreaks\expandafter{\UrlBreaks
  \do\a\do\b\do\c\do\d\do\e\do\f\do\g\do\h\do\i\do\j%
  \do\k\do\l\do\m\do\n\do\o\do\p\do\q\do\r\do\s\do\t%
  \do\u\do\v\do\w\do\x\do\y\do\z\do\A\do\B\do\C\do\D%
  \do\E\do\F\do\G\do\H\do\I\do\J\do\K\do\L\do\M\do\N%
  \do\O\do\P\do\Q\do\R\do\S\do\T\do\U\do\V\do\W\do\X%
  \do\Y\do\Z\do\0\do\1\do\2\do\3\do\4\do\5\do\6\do\7
  \do\8\do\9}

\thicklines
\def\sysname{XRay\xspace}

\def\xray{\sysname}

\ifnum\sandy=1
\fi

\begin{document}

\ifnum\sandy=1
  \doublespacing
\fi
\date{}  % Don't want the date printed.

% \title{\xray: Enhancing the Web's Transparency with Side-Channel Attacks}
% \title{\xray: Enhancing the Web's Transparency with Differential Analysis}
\title{
  \vspace{-40pt}
\xray:  Enhancing the Web's Transparency with Differential Correlation
}
% \title{\xray: Tracking Use of Personal Data on the Web with Differential
% Analysis}
% \title{\xray: Increasing Transparency of Web Services with Differential
% Analysis}
% \title{Tracking Personal Data in the Open Internet with Differential
% Correlation}
%\title{\sysname: Tracking Personal Data in the Open Internet with Stream Correlation}

\ifnum\includeAuthor=1
\author{
{\rm Mathias L\'ecuyer},
{\rm Guillaume Ducoffe},  {\rm Francis Lan},
{\rm Andrei Papancea}, {\rm Theofilos Petsios}, \\ {\rm Riley Spahn},
{\rm Augustin Chaintreau}, and {\rm Roxana Geambasu} \\
    Columbia University
       }
\else
\fi

\maketitle

% Page numbers -- suppress for camera ready.
\ifnum\isCameraReady=1
  \pagestyle{empty}
\fi

% -------------------- %

\begin{abstract}
Today's Web services -- such as Google, Amazon, and Facebook -- leverage user 
data for varied purposes, including personalizing recommendations, targeting
advertisements, and adjusting prices. At present, users have little insight
into how their data is being used. Hence, they cannot make informed choices
about the services they choose.

To increase transparency, we developed {\em \xray}, the first fine-grained,
robust, and scalable personal data tracking system for the Web.  \xray
predicts which data in an arbitrary Web account (such as emails, searches, or
viewed products) is being used to target which outputs (such as ads,
recommended products, or prices). \xray's core functions are service agnostic and
easy to instantiate for new services, and they can track data within and across
services.  To make predictions independent of the audited service, \xray relies
on the following insight:  by comparing outputs from different accounts with
similar, but not identical, subsets of data, one can pinpoint targeting through correlation.
% Constructing a practical tool from this insight raises
% significant unresolved challenges, appearing to require an exponential number of accounts  to pinpoint targeting at fine granularity.
We show both theoretically, and through experiments on Gmail, Amazon, and
YouTube, that \xray achieves high precision and recall by correlating data from
a surprisingly small number of extra accounts.

% Our
% new design surprisingly proves that correlating outputs for multiple inputs can
% successfully pin-point targeting even only with a \emph{logarithmic} number
% of accounts with increasing numbers of audited data items.  We implemented \xray
% on Gmail, Amazon, and YouTube with little porting effort. Our experiments show
% that it tracks data targeting accurately and robustly.

% \xxx{At the end: make contribution clearer + mechanisms more concrete.}
\end{abstract}

\vspace{-8pt}
\section{Introduction}
\label{s:introduction}
\vspace{-8pt}

\iffalse
Recent advances in mobile and cloud technologies are ushering in a new era of
``big data.'' Staggering amounts of personal data -- such as locations,
searches, emails, posts, and photos -- are being collected and analyzed by
Google, Amazon, Facebook, and a myriad of other Web services. This presents rich
opportunities for marshaling big data to improve daily life and social
well-being. Twitter data, for instance, has been successfully applied to public
health problems~\cite{Sadilek:2013gz}, crime prevention~\cite{twitter:2012}, and
emergency response~\cite{Hughes:2009vb}.

Personal data can improve the usability of applications by letting them predict
future user needs and preferences and seamlessly adapt to them.  It can boost
business revenues by enabling effective product placement and targeted
advertisements. Thus, it is of no surprise that the benefits of big data
have generated an across-the-board excitement -- a true frenzy -- with
Web services aggressively pursuing new ways to acquire and use it.
\fi

% Recent advances in mobile and cloud technologies are ushering in a new era of ``big data.''
We live in a ``big data'' world.
Staggering amounts of personal data -- our as locations,
search histories, emails, posts, and photos -- are constantly collected and analyzed by
Google, Amazon, Facebook, and a myriad of other Web services. This presents rich
opportunities for marshaling big data to improve daily life and social
well-being.  For example, personal data improves the usability of
applications by letting them predict and seamlessly adapt to future user needs
and preferences.  It improves business revenues by enabling effective product
placement and targeted advertisements.  Twitter data has been
successfully applied to public health problems~\cite{Sadilek:2013gz}, crime
prevention~\cite{twitter:2012}, and emergency response~\cite{Hughes:2009vb}.
These beneficial uses have generated a big data frenzy, with Web services
aggressively pursuing new ways to acquire and commercialize it.

% Other analogies instead of frenzy/impenetrabele place: data rush/wild west.
Despite its innovative potential, the personal data frenzy has transformed
the Web into an opaque and privacy-insensitive environment.
% unintelligible / impenetrable / lawless land /
Web services accumulate data, exploit it for varied and
undisclosed purposes, retain it for extended periods of time, and possibly
share it with others -- {\em all without the data owner's knowledge or consent}. Who
has what data, and for what purposes is it used? Are the uses in the data
owners' best interests?  Does the service adhere to its own privacy policy? How
long is data used after its owner deletes it?  Who shares data with whom?

At present, users lack answers to these questions, and investigators (such as
FTC agents, journalists, or researchers) lack robust tools to track data in the
ever-changing Web to provide the answers.  Left unchecked, the exciting potential
of big data threatens to become a breeding ground for data abuses, privacy
vulnerabilities, and unfair or deceptive business practices. Examples
of such practices have begun to surface.  In a recent incident, Google was
found to have used institutional emails from ad-free Google Apps for Education to
target ads in users' personal accounts ~\cite{safegov-google, safegov-reply}.
% In doing so, some argue, it is contradicting the general understanding of its
% own privacy policy~\cite{safegov-google}.
MySpace was found to have violated its privacy policy by leaking personally identifiable
information to advertisers~\cite{Krishnamurthy:2009}.
Several consumer sites, such as Orbitz and Staples, were found to have adjusted their product pricing based on
user location~\cite{wsj-orbitz, wsj-staples}. And Facebook's 2010 ad
targeting was shown to be vulnerable to micro-targeted ads specially
crafted to reveal a user's private profile data~\cite{Korolova:2010cq}. % http://theory.stanford.edu/~korolova/
% Privacy_violations_using_microtargeted_ads.pdf
% These are just a few examples of what is likely a broader and growing data
% misuse problem fostered by the Web's obscurity.

To increase transparency and provide checks and balances on data abuse, we
argue that new, robust, and versatile tools are needed to effectively
track the use of personal data on the Web.
%They should be scalable and easily adaptable to a
% wide variety of types of data, Web services, and use cases.
Tracking data in a {\em controlled environment}, such as a modified operating
system, language, or runtime, is an old problem with a well-known solution:
taint tracking systems~\cite{taintdroid, Giffin:2012uo, TaintTrace,
Zhu:EECS-2009-145}. However, is it possible to track data in an {\em
uncontrolled environment}, such as the Web?  Can robust, generic mechanisms
assist in doing so? What kinds of data uses are trackable and what are not?
How would the mechanisms scale with the amount of data being tracked?
% While a few systems exist for tracking the use of a select few data
% types,
% such as IP and search terms~\cite{bobble,other_stuff}, the tools and
% science for tracking the use of arbitrary personal data within Web service
% accounts -- such as emails, documents, or private photos -- are largely
% missing.

As a first step toward answering these questions, we  built {\em \xray},
a personal data tracking system for the Web.
\xray correlates designated data {\em inputs} (be they emails, searches, or visited products)
with data {\em outputs} results (such as ads, recommended products, or prices).
Its correlation mechanism is service agnostic and easy to instantiate, and it can track data
use within and across services.
For example, it lets a data owners track how their emails, Google+, and YouTube
activities are used to target ads in Gmail.

% What we need are robust tools to increase awareness for auditors and end-users
% alike about how the data is used in the data-driven Web.  To this end, we
% present {\em \xray}, a system for tracking personal data use in {\em
% uncontrolled} environments. Core mechanism: differential correlation.

At its core, \xray relies on a {\em differential correlation} mechanism that pinpoints targeting by comparing
outputs in different accounts with similar, but not identical,
subsets of data inputs.  To do so, it associates with every personal
account a number of {\em shadow accounts}, each of which contains different
data subsets.  The correlation mechanism uses a simple Bayesian model to compute
and rank scores for every data input that may have triggered a specific output.
Intuitively, if an ad were seen in many accounts that share a certain email, and
never in accounts that lack that email, then the email is likely to be
responsible for a characteristic that triggers the ad. The email's score for
that ad would therefore be high. Conversely, if the ad were seen rarely in
accounts with or lacking that email, that email's score for this ad would be low.

\iffalse
\F\ref{f:prediction_example} shows a concrete example of personal data use
assessment with differential correlation.  To explain an ad ($ad1$) seen in a
user account with three emails ($e1$, $e2$, and $e3$), \xray inspects the ads
seen in the three shadow accounts associated with the main account.  Because
the ad is seen most often in accounts with email $e2$ (top two shadow accounts)
and never in accounts lacking that email (bottom shadow account), \xray decides
that associating $ad1$ to $e2$ best explains the observations.
\fi

Constructing a practical auditing system around differential correlation raises
significant challenges.  Chief among them is scalability with the number of data items.
Theoretically, \xray requires a shadow account for each combination of data
inputs to accurately pinpoint correlation.  That would suggest an {\em
exponential number of accounts}! Upon closer examination, however, we find that
a few realistic assumptions and novel mechanisms let \xray reach high precision
and recall with only a {\em logarithmic number of accounts in number of data inputs}.
We deem this a major new result for the science of tracking data-targeting
on the Web.
% \xxx{At the end: add more technical discussion and results to intro.}

\iffalse
\begin{figure}[t]
\centering
\includegraphics[width=0.6\linewidth]{figures/prediction_example}
\caption{\small {\bf \xray Correlation Example.}
  \xray yields the ($e2 \rightarrow ad1$) association, which best explains the
observations.
  \label{f:prediction_example}
  \vspace{-10pt}
}
\vspace{-0.2cm}
\end{figure}
\fi

% Although simple, differential correlation is robust, powerful, and versatile.
We built an \xray prototype and used it to correlate Gmail ads, Amazon
product recommendations, and YouTube video suggestions to user emails, wish
lists, and previously watched videos, respectively.
While Amazon and YouTube provide detailed explanations of their targeting, Gmail does not, so we manually validated associations.  For all cases, \xray achieved 80-90\% precision and recall.
Moreover, we integrated our Gmail and YouTube prototypes so we could track cross-service ad targeting.
Although several prior measurement studies~\cite{NikDiak, Xing:2014ws, Hannak:2013uk, Guha:2010hk, Anonymous:2012wi} used methodologies akin to
differential correlation, we believe we are the first to build a generic, service agnostic, and scalable tool based on it.
% Scalability challenge + 3 mechanisms: selective data tracking, data
% clustering, and privacy-preserving collaborative auditing.  Mechanism
% descriptions.
% \xxx{wip} Some other paragraph -- something that raises the reader one level up
% to our greater goal/achievement.
% We reveal data uses for individuals, investigators, and society as a whole to
% judge whether practices are good or bad, fair or unfair, deceptive or
% outright.
Overall, we make the following contributions:
\vspace{-0.1cm}
\begin{enumerate}
\item The first general, versatile, and open system to track arbitrary personal Web data
use by uncontrolled services.  The code is available from our Web page~\url{https://xray.cs.columbia.edu/}.

\item The first in-depth exploration into the scalability challenges of
tracking personal data on the Web.

\item The design and implementation of robust mechanisms to address scaling, including data matching.

\item System instantiation to track data on three services (Gmail,
Amazon, YouTube) and across services (YouTube to Gmail).

\item An evaluation of our system's precision and recall on Gmail, Amazon, and
YouTube.  We show that \xray is accurate and scalable.  Further, it reveals
intriguing practices now in use by Web services and advertisers.

\end{enumerate}
\vspace{-0.3cm}

\iffalse

Concerns about the intro:
1. Diff. correlation seems a fancy name for a known technique. Not clear what
hard problem it solves.
2. Doesn't really make a case for how we're lacking appropriate tools for
tracking on the Web.
3. Major contribution -- scaling -- comes too late and doesn't contain any
detail?
4. Not that exciting?  A few stronger statements needed.  Sandy will likely
help w/ that.
5. Too long.

6. Still unclear what xRay is!!!  Is it a system, is it a set of building
blocks, a framework, what is it???

\fi

\vspace{-4pt}
\section{Motivation}
\label{s:motivation}
\vspace{-8pt}

% A major contribution in this paper is to make the case, and demonstrate a
% promising initial candidate, for a new generation of scalable, robust, and
% versatile tools for increasing data transparency on the Web.  We motivate the
% need for such tools with a set of scenarios (\S\ref{s:scenarios}) and an
% analysis of how current tools fail to address them
% (\S\ref{s:alternative_solutions}).  While our high-level motivation is broad,
% our goals for \xray are more limited and involve a number of simplifying
% assumptions, which we describe in the next section.

This paper lays the algorithmic foundations for a new generation
of scalable, robust, and versatile tools to lift the curtain on how personal data
is being targeted.
We underscore the need for such tools by describing potential usage scenarios
inspired by real-life examples (\S\ref{s:scenarios}).  We do this
not to point fingers at specific service providers; rather, we aim
to show the many situations where transparency tools would be valuable
for end-users and auditors alike.
We conclude this section by briefly analyzing how current approaches fail to address these usage scenarios (\S\ref{s:alternative_solutions}).
% how personal data is often being mined and commercialized in order to illustrate 

\subsection{Usage Scenarios}
\vspace{-3pt}
\label{s:scenarios}

\iffalse
\begin{quote}
{\em ``On a simplistic level, it is true that there are two versions on
  Facebook: the one you obsessively tend, and the hidden, deepest secret in the
  world, which is the data about you that is used to sell access to you to third
  parties like advertisers.  You will never see that second kind of data about
you.''}
~\cite{Jaron_Lanier_Who_Owns_the_Future_p112-114}
\end{quote}
\vspace{-0.2cm}
{\em Or will you?} Let's look at four scenarios to see \xray's value added in
tracing Lanier's ``chain of custody of
data.''~\cite{Jaron_Lanier_Who_Owns_the_Future_p112-114}
\fi

\headingg{Scenario 1: Why This Ad?}  Ann often uses her Gmail ads to
discover new retail offerings.  Recently, she discussed her
ad-clicking practices with her friend Tom, a computer security expert.  Tom
warned her about potential privacy implications of clicking on ads without
knowing what data they target.  For example, if she clicks on an ad targeting
the keyword ``gay'' and then authenticates to purchase something from that
vendor, she is unwittingly volunteering potentially sensitive information to the
vendor.
% Given how little information Gmail reveals about ad targeting,
Tom tells Ann about two options to protect her privacy.  She can either disable the ads
altogether (using a system like AdBlock~\cite{adblock}), or install the \xray
Gmail plugin to uncover targeting against her data.  Unwilling to give up
the convenience of ads, Ann chooses the latter.  \xray clearly annotates the ads in the Gmail UI with their target email or combination, if any.
Ann now inspects this targeting before clicking on an ad and avoids clicking if highly sensitive emails are being targeted.

% Ann, a regular user, likes
% her Gmail ads, which she ﬁnds highly relevant and
% convenient. She has found some of her most favorite
% clothes through such ads. Recently, however, she has
% been reﬂecting on the privacy implications of using the
% ads without knowing what they are targeted against.
% What information is she unwittingly volunteering to a
% vendor by clicking on their ad and then authenticating to
% perform a purchase? What if a vendor had targeted the
% ad toward the keyword ``gay?'' Does this mean that the
% vendor now suspects that she is gay?
%
% Ann tries to use the ``Why This Ad?'' link that
% Google displays for every ad, but the information is
% disappointingly vague: ``These ads are based on emails
% from your mailbox and information from your Google
% account.'' Recently, she has installed the xRay Gmail
% plugin, which shows her in detail which emails have
% likely triggered which ads. Now, before clicking on an
% ad, Ann consults \xray's predictions (shown in Figure 1)
% to ensure that the ad is not targeted against something
% that she would be uncomfortable sharing with the target
% website. In truth, Ann would have preferred – and
% expected – Google to have been more transparent so she
% needn’t install a third-party tool.

\heading{Scenario 2: They're Targeting {\em What}?}
% \footnote{Scenario 2 is
% inspired from our \xray experience, described in \S\ref{s:eval:experience}.}
Bob, an FTC investigator, uses the \xray Gmail plugin for a different purpose:
to study sensitive-data targeting practices by advertisers. He suspects a
potentially unfair practice whereby companies use Google's ad network to
collect sensitive information about their customers. Therefore, Bob creates a
number of emails containing keywords such as ``cancer,'' ``AIDS,''
``bankruptcy,'' and ``unemployment.'' He refreshes the Gmail page many
times, each time recording the targeted ads and \xray's explanations for them.
The experiment reveals an interesting result: an online insurance company,
TrustInUs.com, has targeted multiple ads against his illness-related emails.
Bob hypothesizes that the company might use the data to set higher premiums for
users reaching their site through a disease-targeted ad.  He uses \xray
results as initial evidence to open an investigation of TrustInUs.com.

\heading{Scenario 3: What's With The New Policy?}\footnote{In Feb. 2014, it
was revealed based on court documents that Google could have used institutional
emails to target ads in personal accounts~\cite{safegov-google}.
In May 2014, Google committed to disable that
feature~\cite{google-promise-apps-for-ed}. Scenario 3 presents an \xray-based approach
to investigate the original allegation.}
Carla, an investigative journalist, has set up a watcher on privacy policies for major Web services.
When a change occurs, the watcher notifies her of the difference.
Recently, an important sentence in Google's privacy policy has been scrapped:
\vspace{-0.2cm}
\begin{quote}
{\em
If you are using Google Apps (free edition), email is scanned so we can
display conceptually relevant advertising in some circumstances. \sout{Note
that there is no ad-related scanning or processing in Google Apps for Education
or Business with ads disabled.}
}
\end{quote}
\vspace{-0.2cm}
To investigate scientifically whether this omission represents a shift in
implemented policy, she obtains institutional accounts, connects them to
personal accounts, and uses \xray to detect
the correlation between emails in institutional accounts and ads
in corresponding personal accounts.  Finding a
strong correlation, Carla writes an article to expose the policy change and
its implications.

\heading{Scenario 4: Does Delete Mean Delete?}  Dan, a CS researcher, has seen
the latest news that Snapchat, an ephemeral-image sharing Website, does not
destroy users' images after the requested timeout but instead just unlinks
them~\cite{Snapc0:Online}.  He wonders whether the reasons for this are purely
technical as the company has declared (e.g., flash wearing levels, undelete
support, spam filtering)~\cite{snapchatblog, snapchatblog2} or whether these
photos, or metadata drawn from them, are mined to target ads or other products
on the Website.  The answer will influence his decision about whether to
continue using the service. Dan instantiates \xray to track the correlation
between his expired Snapchat photos and ads.
% \xxx[mathias]{Explain what he needs to do in 2 lines.}

\iffalse
\heading{Example 5: Are They Selling User Data?}  \xxx{Enabled for
Sandy's consideration -- by default disable.}  Bob, an FTC investigator,
has received some tips about questionable data exchanges between Websites
\url{a.com} and \url{b.com}.  To gain sufficient evidence for a full legal
investigation, he customizes \xray to track the correlation between the data
in his \url{a.com} account and the outputs in his \url{b.com} account. For
this, he \xxx[mathias]{what does he need to do, in one sentence}.  \xray
suggests strong correlation between some outputs and some inputs.
Bob concludes that the tip is worthy of further investigation and uses the
correlation evidence to pursue court warrants necessary to establish
causality.
\fi

\subsection{Alternative Approaches}
\vspace{-3pt}
\label{s:alternative_solutions}

The preceding scenarios illustrate the importance of transparency in
protecting privacy across a range of use cases.
% % Examples 1 and 4 illustrate how providing end-users with
% % information about how their data is being used helps them calibrate their
% % privacy expectations and make more informed decisions about the services they
% % use, and how they use them.  Examples 2 and 3 illustrate how journalists and
% % investigators can leverage \xray as a building block for their investigations.
% % Moreover, the kinds of data, outputs, and services to audit differ depending
% % on the use case.
{\em We need robust, generic auditing tools to track the use of personal data
at fine granularity (e.g., individual emails, photos) within and across
 arbitrary Web services.}  At present, no such tools exist, and the science of
tracking the use of personal Web data at a fine grain is largely non-existent.

Existing approaches can be broadly classified in two categories:
{\em protection tools}, which prevent Web services' acquisition or use of personal
data, and (2) {\em auditing tools}, which uncover
Web services' acquisition or use of personal data.
We discuss these approaches next; further related work is in \S\ref{s:relwork}.

\heading{Protection Tools.}
A variety of protection tools exist~\cite{Dingledine:2004tp,
Roesner:2012uj, adblock, pgp}. For example, Ann could disable ads using an ad
blocker~\cite{adblock}. Alternatively, she could encrypt her emails,
particularly the sensitive ones, to prevent Google from using them to target
ads.  Dan could use a self-destructing data system, such as
Vanish~\cite{vanish}, to ensure the ephemerality of his Snapchat photos.

While we encourage the use of protection tools, they impose
difficult tradeoffs that make them inapplicable in many cases.  If Ann blocks
all her ads, she cannot benefit from those she might find
useful; if she encrypts all of her emails, she cannot search
them;  if she encrypts only her sensitive emails, she cannot protect
any sensitive emails she neglected to encrypt in advance.  Similarly, if Dan
encrypts his Snapchat photos, sharing them becomes more difficult.  While more
sophisticated protection systems address certain limitations (e.g.,
searchable~\cite{peks}, homomorphic~\cite{fhe, cryptdb}, and attribute-based
encryption~\cite{attribute_based_encryption_for_fine_grained}, or
privacy-preserving advertising~\cite{Toubiana:2010tm,
Fredrikson:2011dn}), they are generally heavyweight~\cite{fhe}, difficult to
use~\cite{Whitten:1999tu}, or require major service-side changes~\cite{fhe, Toubiana:2010tm, Fredrikson:2011dn}.
% Accordingly, these systems have yet to see widespread adoption.

\heading{Auditing Tools.}
Given the limitations of protection tools, transparency is gaining increased
attention~\cite{Xing:2014ws, taintdroid, Hannak:2013uk}.  If protecting data
proves too cumbersome, limiting, or unsupportive of business needs, then users
should at least be able to know: (1) {\em who is handling their data?}, and (2)
{\em what is it being used for?}
% This is the view we take with \xray.

Several tools developed in recent years partially address the
first question by revealing where personal data flows from a local
device~\cite{sharemenot, taintdroid, collusion}.
% Those can be used either by the end-users or by third-party investigators.
TaintDroid~\cite{taintdroid} uses taint tracking to detect leakage of personal
data from a mobile application to a service or third-party backend.
ShareMeNot~\cite{sharemenot} and Mozilla's Lightbeam Firefox add-on~\cite{lightbeam} identify
third parties that are observing user activities across the Web.
These systems track personal data -- such as location, sensor data, Web
searches, or visited sites -- {\em until it leaves the user's
device}. Once the data is uploaded to Web services,  it
can be used or sold without a trace.  In
contrast, \xray's tracking just begins: we aim to tell users how services use their data {\em once they have it}.

Several new tools and personalization measurement studies partially address the
second question: what data is being used for~\cite{NikDiak, Xing:2014ws,
Hannak:2013uk, Guha:2010hk, Anonymous:2012wi}.
In general, all existing tools are highly specialized, focusing on specific input types, outputs, or services.
No general, principled foundation for data use auditing exists, that can be
applied effectively to many services, a primary motivation for this our work.
For example, Bobble~\cite{Xing:2014ws} reveals search result
personalization based on user location (\eg IP) and search history.
% Its basic functioning resembles differential correlation -- detecting personalization by comparing outputs from different profiles -- a notion that we did not invent, but for which we are laying the scientific foundations.
Moreover, existing tools aim to discover only {\em whether} certain
types of user inputs -- such as search history, browsing history, IP, etc. --
influence the output.  None pinpoints at fine grain {\em which}
specific input -- which search query, which visited site, or which viewed
product -- or combination of inputs explain which output.
\xray, whose goals we describe next, aims to do just that.

\vspace{-8pt}
\section{Goals and Models}
\vspace{-8pt}
\label{s:goals_assumptions}

Our overarching goal is to develop the core abstractions and
mechanisms for tracking data within and across arbitrary Web sites.  After describing specific goals (\S\ref{s:goals}), we narrow our scope
with a set of simplifying assumptions regarding the data uses that
\xray is designed to audit (\S\ref{s:service_model}) and the threats
it addresses (\S\ref{s:threat_model}).

\subsection{Goals}
\vspace{-3pt}
\label{s:goals}

Three specific goals have guided \sysname's design:

\headingi{{\bf Goal 1:} Fine-Grained and Accurate Data Tracking.}  Detect which
specific data inputs (e.g., emails) have likely triggered a particular output
(e.g., an ad).  While coarse-grained data use information (such as Gmail's
typical statement, ``This ad is based on emails from your
mailbox.'') may suffice at times, knowing the specifics can be revelatory,
particularly when the input is highly sensitive and aggressively targeted.

\headingi{{\bf Goal 2:} Scalability.}  Make it practical
to track significant amounts of data (e.g., past month's emails).
We aim to support the tracking of hundreds of inputs
with reasonable costs in terms of shadow accounts.  These accounts are
generally scarce resource since their creation is being constrained
by Web services.   While we assume that users and auditors can obtain
{\em some} accounts on the Web services they audit (e.g., a couple dozen),
we strive to minimize the number required for accurate and
fine-grained data tracking.

\headingi{{\bf Goal 3:} Extensibility, Generality, and Self-Tuning.}
Make \xray generic and easy to instantiate for many services and input/output types.
Instantiating \xray to track data on new Web sites should be simple,
although it may require some service-specific implementation of input/output
monitoring.  However, \xray's correlation machinery -- the conceptually
challenging part of a scalable auditing tool -- should be turn key
and require no manual tuning.
% This forces us to make few assumptions about the services and to rely on
% robust, self-tuning mechanisms to track data.
% For example, a potential
% approach to track, say, email-to-ad relations could textually match the
% emails to the ads based on common keywords, for example.
% We decided that such techniques were too service specific, trying to
% approximate its functioning.

% On their face, these goals appear a fool's dream for several reasons.
% First, the Web, an extremely heterogeneous environment, has as many
% personal data uses as services.  How will we abstract diversity away to
% design robust and generic building blocks for data tracking?  Second, not all
% uses of personal data are auditable from a third-party's
% vantage point.  For example, a company may review user data internally to
% tune a particular feature; this use is invisible to an outsider.
% Third, the space of possible data inputs and outputs is gigantic. How will the tracking mechanisms scale with so many inputs and outputs?

\subsection{Web Service Model}
\vspace{-3pt}
\label{s:service_model}

These goals may appear unsurmountable. An extremely
heterogeneous environment, the Web has perhaps as many data uses as services.
Moreover, data mining algorithms can be
complex and proprietary.
How can we abstract away this diversity and complexity to design robust and
generic building blocks for scalable data tracking?
Fortunately, we find that certain popular classes of Web data uses
lend themselves to principled abstractions that facilitate scalable
tracking.

\begin{figure}[t]
\centering
\includegraphics[width=.78\linewidth]{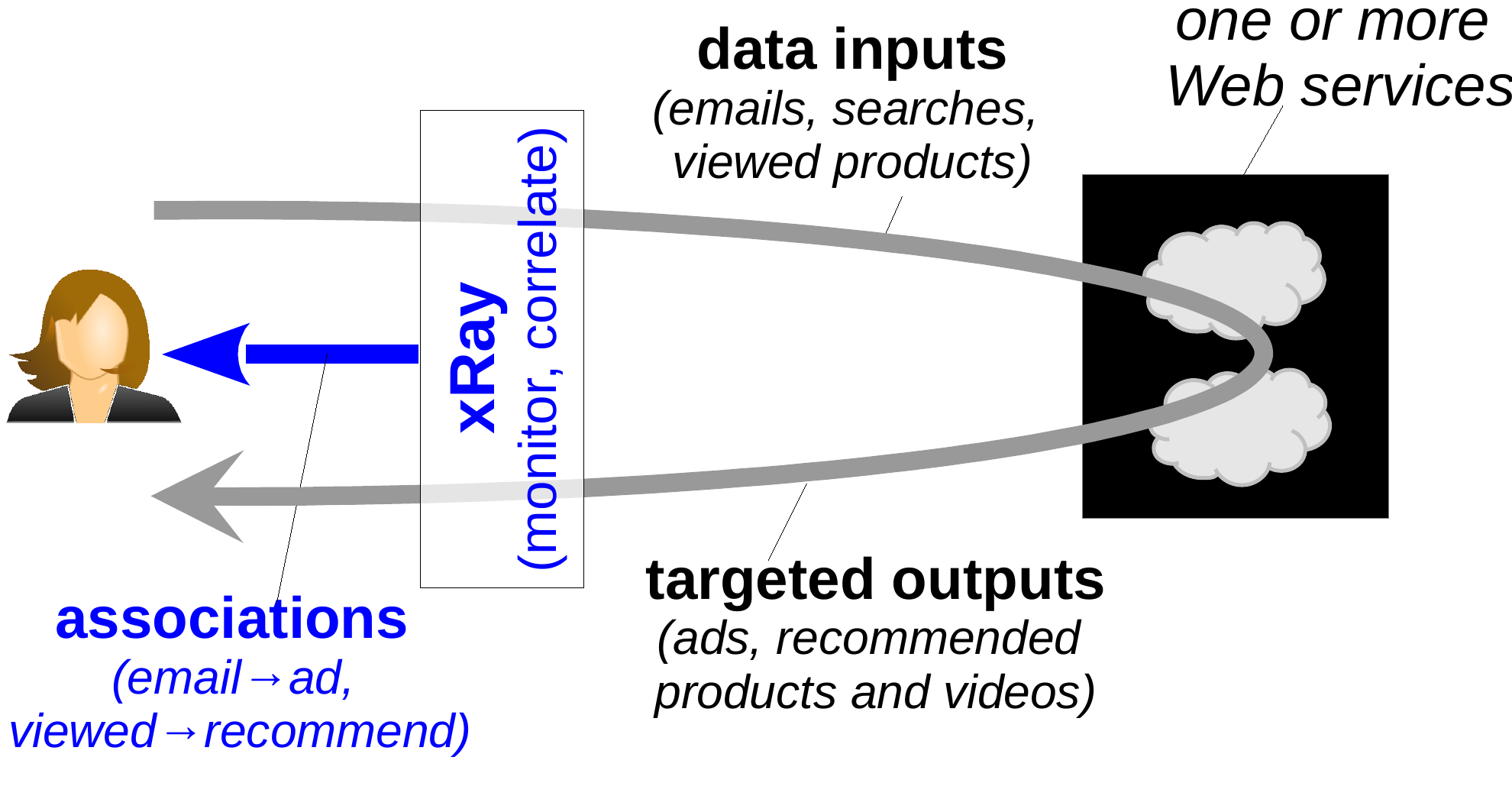}
\vspace{-0.2cm}
\caption{\small {\bf \xray Conceptual View.}  \xray views Web services as
  black boxes, monitors user {\em inputs} and {\em outputs} to/from them, and
  detects  data use through correlation.  It returns to the user or auditor {\em
  associations} of specific inputs and outputs.
  \vspace{-0.5cm}
}
\label{f:inputs_outputs_view}
\end{figure}

\F\ref{f:inputs_outputs_view} shows \xray's simplified view of Web services.
Services, and networks of services that exchange user data, are {\em black boxes}
that receive personal data {\em inputs} from users -- such as emails,
pictures, search queries, locations, or purchases -- and use them for varied
purposes.  Some uses materialize into {\em outputs} visible to
users, such as ads, product or video recommendations, or prices.
Others invisible to the users.  \xray correlates some visible data inputs
with some visible outputs by monitoring them,
correlating them, and reporting strong {\em associations} to users.
An example association is which email(s) contributed to the selection of a particular ad.

\xray relates only {\em strongly correlated} inputs with outputs.  If an
output is strongly correlated to an input (i.e., the input's presence or
absence changes the output), then \xray will likely be able to detect its use.
If not (i.e., the monitored input plays but a small
role in the output), then it may go undetected.
\xray also relates small combinations of inputs with strongly correlated outputs.

Although simple, this model efficiently addresses several
types of personal data functions, including product recommendations, price
discriminations, and various personalization functions (e.g., search, news).
We refer to such functions generically as {\em targeting functions} and focus
\xray's design on them.

Three popular forms of targeting are:  %~\cite{cite-something-official}:
\begin{enumerate}

\item {\em Profile Targeting}, which leverages static or slowly evolving explicit
information -- such as age, gender, race, or location -- that the user often supplies
by filling a form.  This type of targeting has been studied profusely
\cite{NikDiak, Xing:2014ws, Hannak:2013uk, Guha:2010hk, Anonymous:2012wi}; we thus ignore it here.

\item {\em Contextual Targeting}, which leverages the content currently being displayed. In
Gmail, this is the currently open email next to which the ad is shown.
In Amazon or Youtube, the target is the product or video next to which the recommendation is shown.

\item {\em Behavioral Targeting}, which leverages a user's past actions.  An
email sent or received today can trigger an ad tomorrow; a video watched now can
trigger a recommendation later.  Use of histories makes it harder for users to
track which data is being used, a key motivation for our development of \xray.

\end{enumerate}

Theoretically, our differential correlation algorithms could be applied
to all three forms of targeting. From a systems perspective, \xray's design is geared towards {\em contextual targeting} and {\em a specific form of behavioral targeting}.  The latter requires further attention.   We observe that this broad targeting class subsumes multiple types of targeting that operate at different granularities.
For example, a service could use as inputs a user's most recent few emails to
decide targeting. This would be similar to an extended context.
Alternatively, a service could use historical input to learn a user's coarse interests or characteristics and
base its targeting on that.

\xray currently aims to disclose any targeting applied at the level of individual user data, or small combinations thereof.
Our differential correlation algorithms could be applied to
detect targeting that operates on a coarser granularity.
However, the \xray system itself would require significant changes.
Unless otherwise noted, we use {\em behavioral targeting} to denote the restricted
form of behavioral targeting that \xray is designed to address.
We formalize these restrictions in \S\ref{s:formalized_model}.

% Finally, the described Web service model says nothing about which services
% receive the inputs and produce the outputs.  Indeed, the model can be applied to
% relate inputs within one service to outputs within a different one, thereby
% innately supporting cross-service tracking to reveal data sharing on the Web.

\subsection{Threat Model}
\vspace{-3pt}
\label{s:threat_model}

To further narrow our problem's scope, even further, we introduce threat
assumptions.  We assume that data owners (users and auditors) are trusted and
do not attempt to leverage \xray to harm Web services or the Web
ecosystem. While they trust Web services with their data, they wish to better
understand how that data is being used. Data owners are thus assumed to upload
the data in cleartext to the Web services.
% and to make no efforts to thwart
% the services' data collection.

% Web services aim to leverage users' data to target various outputs at them,
% such as ads, product recommendations, etc. They do so for financial purposes
% or to improve the general functioning of their applications.
The threat models relevant for Web services depend on the use case.  For
example, Scenarios 1 and 2 in \S\ref{s:scenarios} assume Google is trusted, but its users wish to understand more about how advertisers target
them through its ad platform.  In contrast, in Scenarios 3 and 4,
investigators may have reason to believe that Web services might
intentionally frustrate auditing.

This paper assumes an {\em honest-but-curious} model for Web services: they
try to use private data for financial or functional gains, but they
do not try to frustrate our auditing mechanism, e.g., by identifying and disabling shadow
accounts.  The service might attempt to
defend itself against more general types of attacks, such as spammers or DDoS
attacks.  For example, many Web services constrain the creation of
accounts so as to limit spamming and false clicks.  Similarly, Web
services may rate limit or block the IPs of aggressive data collectors.  \xray
must be robust to such inherent defenses.
% Moreover, we assume that third-party
% advertisers are untrusted and may try to hide their targeting from \xray.
% These assumptions let us focus on other core challenges of building a robust
% auditing system, such as making it scale with the number of inputs.
We discuss challenges and potential approaches for stronger adversarial models in
\S\ref{s:security_analysis}.

\vspace{-8pt}
\section{The \xray Architecture}
\vspace{-8pt}
\label{s:architecture}

% Shorter
\xray's design addresses the preceding goals and assumptions. For
concreteness, we draw examples from our three \xray instantiations: tracking
email-to-ad targeting association within Gmail, attributing recommended videos
to those already seen on YouTube, and identifying products in a
wish list that generate a recommendation on Amazon.

%We next describe our \xray design, which addresses the preceding goals.  Although \xray is generally applicable to many Web services and use cases, we use examples from our Gmail, YouTube, and Amazon experience.  On Gmail, we track email-to-ad associations; on YouTube, we explain why certain videos are recommended to users based on the videos they viewed previously; and on Amazon, we explain which products are recommended to users based on the contents of their carts and wish lists.

\subsection{Architectural Overview}
\vspace{-3pt}
\label{s:overview}

\begin{figure}[t]
\centering
\includegraphics[width=\linewidth]{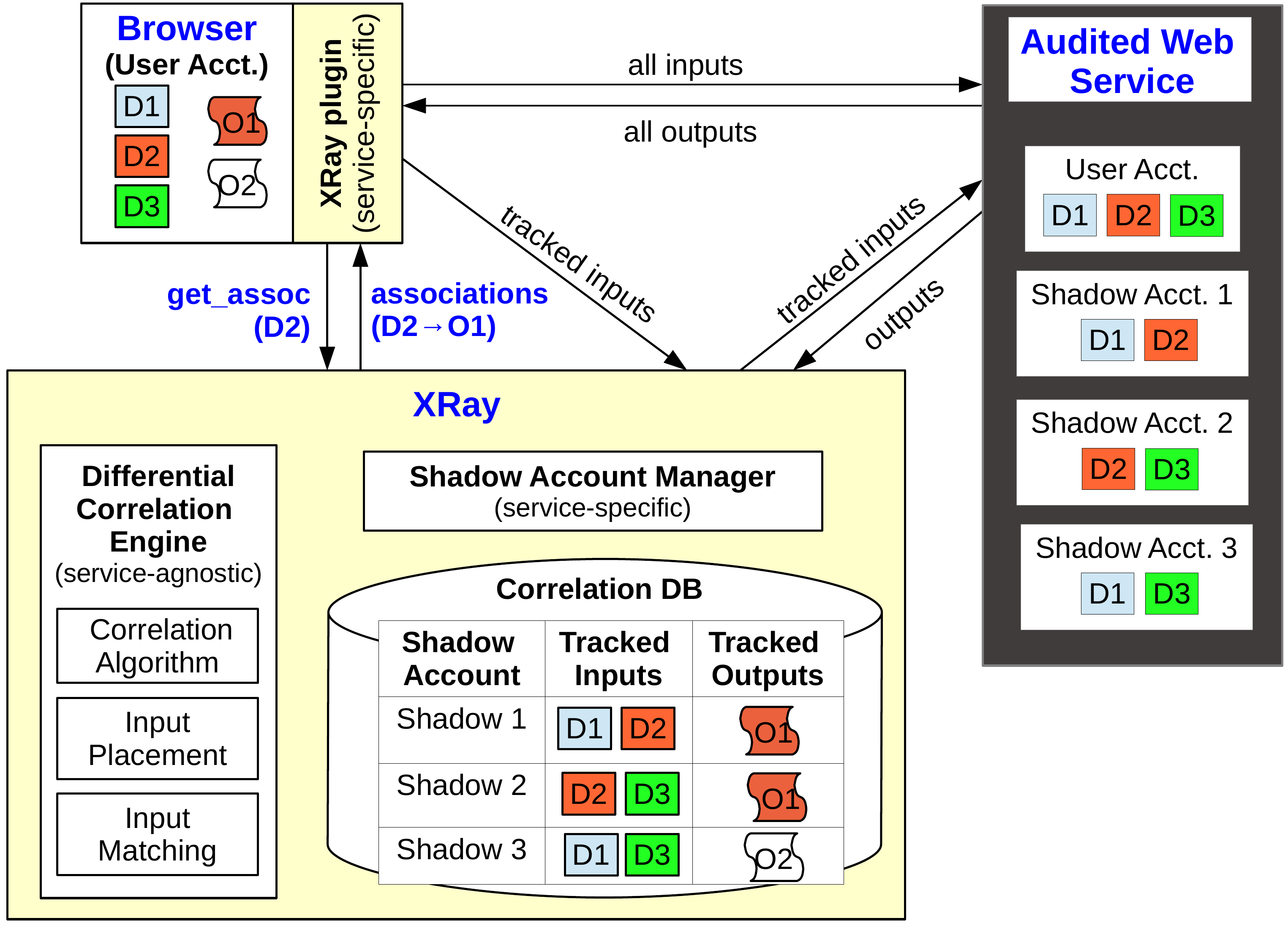}
\caption{\small {\bf The \xray Architecture.}
}
\label{f:architecture}
\vspace{-0.1cm}
\end{figure}

\xray's high-level architecture (\F\ref{f:architecture}) consists of
three components: (1) a {\em Browser Plugin}, which intercepts tracked inputs
and outputs to/from an audited Web service and gives users visual
feedback about any input/output associations, (2) a {\em Shadow Account
Manager}, which populates shadow accounts with inputs from the plugin and
collects outputs (e.g., ads) for each shadow account, and (3) the {\em
Correlation Engine}, \xray's core, which infers associations and provides them to
the plugin for visualization.  While the Browser Plugin and Shadow Account
Manager are {\em service specific}, the Correlation Engine, which encapsulates
the science of Web-data tracking, is {\em service agnostic}.
% For example,
% implementing the plugin and shadow account managers for Gmail, Amazon, and
% YouTube was conceptually simple and only took about 500 lines of code per
% service; zero changes were needed in the correlation engine.
After we describe each component, we focus on the design of the
Correlation Engine.

\heading{Browser Plugin}.  The Browser Plugin intercepts designated inputs and outputs (i.e., {\em tracked
inputs/outputs}) by recognizing specific DOM elements in an audited service's
Web pages.  Other inputs and outputs may not be tracked by \xray
(i.e., {\em untracked inputs/outputs}).  The decision of what to track belongs to an investigator or developer who instantiates
\xray to work on a specific service. For example, we configure the \xray Gmail Plugin
to monitor a user's emails as inputs and ads as outputs.
When the Plugin gets a new tracked input (e.g., a new email), it forwards it both
to the service and to the Shadow Account Manager.
When the Plugin gets a new tracked output (e.g., an ad), it queries the
Correlation Engine for associations with the user's tracked inputs (message
{\code get\_assoc}).
% It displays any associations returned to the user.
% \F\ref{f:screenshot} shows a specific visualization of associations
% that we implemented for our Gmail \xray plugin.  It shows above each ad a
% colored label indicating whether the ad is targeted; clicking on a targeted
% label shows which email(s) triggered the ad.

\heading{Shadow Account Manager.}  This component: (1) populates the shadow
accounts with subsets of a user account's tracked inputs (denoted $D_i$),
and (2) periodically retrieves outputs (denoted $O_k$) from the audited service
for each shadow account.  Both
functions are service specific.  For Gmail, they send emails with
SMTP and call the ad API.  For YouTube, they
stream a video and scrape recommendations, and for Amazon, they
place products in wish lists and scrape recommendations.  The complexity of
these tasks depends on the availability of APIs or the stability of a service's
page formats.  Outputs collected from the Web service
are placed into a {\em Correlation Database} (DB), which maps shadow accounts to their
input sets and output observations.
\F\ref{f:architecture} shows a particular assignment of tracked
inputs across three shadow accounts. For example, Shadow 1 has inputs $D_1$ and
$D_2$.  The figure also shows the outputs collected for each shadow
account.  Output $O_1$ appears in Shadows 1 and 2 but not in 3; output
$O_2$ appears in Shadow 3 only.

\heading{Differential Correlation Engine.}
This engine, \xray's service-agnostic ``brain,''
leverages the data collected in the Correlation DB to infer input/output
associations.
% The Correlation Engine consists of three components: {\em Correlation Algorithm},
% {\em Input Placement}, and {\em Input Matching}, which we next overview.
% Three key functions ensure that associations can be best inferred under
% stringent shadow account constraints.  First, the {\em Correlation Algorithms}
% module performs Bayesian inference to detect associations,
% incrementally leveraging new output observations. Second, the {\em Input
% Placement} module distributes tracked inputs across shadow accounts.
% Third, the {\em Input Matching} module lets \xray use shadow accounts to maximum
% advantage by reducing overlaps between distributed inputs, which creates noise
% for \xray.
When new outputs from
shadow accounts are added into the Correlation DB, the engine
attempts to diagnose them using a {\em Correlation Algorithm}.
We developed several such algorithms and describe them in \S\ref{s:correlation_algos}.
% We have developed several algorithms and describe them in subsequent sections.
This process, potentially time-consuming process, is done as a background job,
asynchronously from any user request.
In \F\ref{f:architecture}, differential correlation might conclude that $D_2$
triggers $O_1$ because $O_1$ appears consistently in accounts with that $D_2$.
It might also conclude that $O_2$ is {\em untargeted} given inconsistent
observations.  The engine saves these associations in the Correlation DB.

When the plugin makes a {\code get\_assoc} request, the Correlation Engine
looks up the specified output in its DB and returns any pre-computed association.
If no output is found, then the engine replies  {\em unknown} (e.g.,
if an ad never appeared in any shadow account or there is
insufficient information).
Periodic data collection, coupled with an online update of correlation model parameters, minimizes the number of unknown associations.
Our experience shows that collecting shadow account outputs in Gmail every ten hours
or so yielded few unknown ads.

While the preceding example is simple, \xray can handle complex challenges
occurring in practice.  First, outputs are never consistently
seen across all shadow accounts containing the input they target. We call
this the {\em limited-coverage} problem; \xray handles it by placing each data input in more shadow accounts.
% Perhaps surprisingly, we prove that under realistic
% assumptions, the required number of accounts grows logarithmically in the number
% of inputs, where the log's constant depends on the coverage (low coverage means
% a higher constant).  \S\ref{s:eval:scalability} validates this theoretical
% result with experiments on Gmail, Amazon, and YouTube.
Second, an output may have been triggered by one of several targeted inputs (\eg
multiple emails on the same topic may cause related ads to appear), a problem we
refer to as \emph{overlapping-inputs}. This exacerbates the number of
accounts needed, since it diminishes the differential signal we receive
from them.
\xray uses robust, service-agnostic mechanisms and algorithms to match
overlapping inputs, place them in the same accounts, and detects their use as a
group.
% Grouping inputs \emph{a priori} seems to contradict our stated goal
% not to depend on their content. Remarkably, \xray performs {\em input matching},
% a technique that leverages partial observations from the outputs
% to matching overlapping inputs. This boosts differential signals and hence \xray's efficiency.

\heading{Organization.}
The remainder of this section describes the Differential Correlation Engine.
After constructing it for Gmail, we applied it as-is for Amazon and YouTube,
where it achieved equally high accuracy and scalability despite
observable differences in how targeting works on these three services.
After establishing notations and formalizing our assumptions
(\S\ref{s:formalized_model}), we describe multiple correlation algorithms,
which build up to our self-tuning correlation algorithm that made this adaptation
convenient (\S\ref{s:correlation_algos}). \S\ref{s:input_matching} describes
our input matching.

\subsection{Notation and Assumptions}
\vspace{-3pt}
\label{s:formalized_model}

We use $f$ to denote the black-box function that represents the
service (\eg Gmail) associating inputs $D_i$s (\eg the emails received and sent)
to targeted outputs $O_k$s (\eg ads).  Other inputs are either ignored by \xray,
known only to the targeting system, or under no known control.  We assume they
are independent or fixed, captured in the randomness of $f$.

We assume that $f$ decides targeting using: (1) a single input (\eg show $O_k$ if
$D_4$ is in the account), (2) a conjunctive combination of inputs (\eg show $O_k$ if
$D_5$ {\em and} $D_8$ are in the account), or (3) a disjunctive combination of the
previous (\eg show $O_k$ if ($D_5$ {\em and} $D_8$) are in the account {\em or}
if $D_4$ is in the account).  We refer to conjunctive and disjunctive
combinations as AND and OR combinations, respectively,
and assume that their is bounded by a maximum {\em input size}, $r$.
This corresponds to the preceding definition of behavioral targeting from \S\ref{s:service_model}.
Contextual targeting will always be a single-input (size-one) combination.

Our goal is to decide whether $f$ produced each output $O_k$
as a reaction to a bounded-size combination of the $D_i$s.
We define as {\em untargeted} any ad that is not targeted against any
combination of $D_i$s, though in reality the ad could be targeted against
untracked inputs.  We denote untargeting as $D_\emptyset$, meaning that the ad is targeted against the ``void'' email.
Our algorithms compute the most likely combination from the $N$ inputs that
explains a particular set of observations, $\vec{x}$, obtained by \xray.

We define three probabilities upon which our algorithms and analyses depend.
First, the {\em coverage}, $\pin$, is the probability that an account $j$ containing the input $D_i$ targeted by a particular ad, will see that ad at least once.
Second, an account $j'$ lacking input $D_i$ will see the ad with a smaller probability, $\pout$.
Third, if the ad is not behaviorally targeted, it will appear in each account with the same probability, $\punt$.
We assume that $\pin,\punt,\pout$ are constant across all emails, ads, and time, and that $\pout$ is strictly smaller than $\pin$ (bounded noise hypothesis).

% We also assume that every targeting can be explained by a
% combination of inputs of bounded size, and that the targeting function is
% monotonous (adding inputs only add relevant outputs: there is no negative
% targeting).
Finally, we consider all outputs to be independent of each other across time.
\S\ref{s:discussion} discusses the implications.

\subsection{Correlation Algorithms}
\vspace{-3pt}
\label{s:correlation_algos}

A core contribution of this paper is our service-agnostic, self-tuning
differential correlation algorithm, which requires only a logarithmic
number of shadow accounts to achieve high accuracy.  We wished not
only to validate this result experimentally, but also to prove it theoretically
in the context of our assumptions.  This section constructs
the algorithm in steps, starting with a na\"ive polynomial algorithm
that illustrates the scaling challenges.
We then define a base algorithm using set intersections and
prove that it has the desired logarithmic scaling properties; it has
parameters which, if not carefully chosen, can lead to poor
results.
We therefore extend this base algorithm into a self-tuning Bayesian model
that automatically adjusts its parameters to maximize correctness.

\subsubsection{Na\"ive Non-Logarithmic Algorithm}
\vspace{-3pt}
\label{s:naive}

An intuitive approach to differential correlation is to create accounts for
every combination of inputs, gathering maximum information about their
behaviors. With a sufficient number of observations, one could expect
to detect which accounts, and hence which subsets of
inputs, target a particular ad. Unfortunately, this method requires a number of
accounts that grows \emph{exponentially} as the number of items $N$ to track
grows.
When restricting the size of combinations to $r$, as we do in \xray,
the number of accounts needed
is {\em polynomial} (in $O(N^r)$), or {\em linear} if we study
unique inputs only.
Even a linear number of accounts in the number $N$ of inputs remains
impractical to scale to large input sizes (e.g., a mailbox).

\subsubsection{Threshold Set Intersection}
\vspace{-3pt}
\label{s:setintersection}

We now show that it is possible to infer behavioral
targeting using no more than a \emph{logarithmic} number of accounts as a
function of the number of inputs. Specifically, we prove the following
theorem:

\begin{thm}\label{thm:set-intersection}
  Under \S\ref{s:formalized_model} assumptions, for any $\varepsilon > 0$ there exists an algorithm that requires $C \times \ln(N)$ accounts to correctly identify the inputs of a targeted ad with probability $(1-\varepsilon)$.
  The constant C depends on $\varepsilon$ and the maximum size
  of combinations $r$ ($O(r 2^r\log(\frac{1}{\varepsilon}))$).
  \label{res:placement}
\end{thm}

% The key insight is that combining multiple inputs inside one account provides
% extra information that can be combined to inform the inference. Intuitively, a
% logarithmic number of observations provides $\log(N)$ bits, which is the
% appropriate order to choose correctly an element in a set of $N$ hypotheses.

To demonstrate the theorem, we define the {\em Set Intersection Algorithm} and prove that it has the correctness and scaling properties specified in the theorem.  Given that outputs will appear more often in accounts containing the targeting inputs, the core of the algorithm is to determine the set of inputs appearing in the highest number of accounts that also see a given ad.
This paper describes a basic version of the algorithm that makes some simplifying
assumptions and provides a brief proof sketch.  
\ifnum\isTR=1
The detailed proof and
complete algorithm are described in Appendix.
\else
The detailed proof and
complete algorithm are described in our technical report~\cite{xray-tr}.
\fi

\begin{figure}[t]
\centering
\begin{minipage}{0.92\linewidth}
\begin{lstlisting}[mathescape]
// `\textbf{Set Intersection Algo:}`
// Runs with each collected ad.
`\textbf{In}`: Output $O_k$ (e.g. an ad).
`\textbf{Params}`: MIN_ACTIVE_ACCTS, THRESHOLD.
`\textbf{Out}`: Targeted input combination.
// Step 1: Compute active accounts.
$\aacct$ = the accounts that see ad $O_k$.
if $|\aacct|$ < MIN_ACTIVE_ACCTS
  return $\emptyset$
end
// Step 2: Create input combination hypothesis.
targeted_set = $\emptyset$
`{\bf foreach}` input $D_i$ do
  if $\frac{number~of~\aacct~containing~D_i}{|\aacct|} > $THRESHOLD
    targeted_set += $D_i$
  end
end
// Step 3: Verify it is a real combination.
if $\frac{number~of~\aacct~containing~entire~targeted\_set}{|\aacct|} < $THRESHOLD
  return $\emptyset$
end
// targeted_set triggered the output.
return targeted_set
\end{lstlisting}
\end{minipage}
\vspace{-0.4cm}
\caption{{\small {\bf The Set Intersection Algorithm.} Can be proven to predict targeting correctly under certain assumptions with a logarithmic number of accounts.}}
\label{f:algo}
\end{figure}

\heading{Algorithm.}
The algorithm relies on a randomized placement of inputs into shadow
accounts, with some redundancy to cope with imperfect coverage.  We thus
pick a probability, $0<\alpha<1$, create $C \ln(N)$ shadow accounts,
and place each input $D_i$ randomly into each account with probability $\alpha$.
\F\ref{f:algo} shows the Set Intersection algorithm for a set of
observations, $\vec{x}$.
Given an output $O_k$ collected from the user account, we
compute the set of {\em active accounts}, $\aacct$, as those shadow
accounts that have seen the output (Step 1).
We then compute the set of inputs that
appear in at least a threshold fraction of active accounts;
this set is our candidate for the combination being targeted by the ad
(Step 2).
Finally, we check that the entire combination is in a threshold fraction of
the active accounts (Step 3).
% \footnote{The thresholds in Steps 2 and 3 can be
% different or the same, the proof does not change.}
Theoretically, we prove that there
exists a threshold for which the algorithm is arbitrarily correct with
the available $C \ln(N)$ accounts.
Practically, this threshold must be tuned experimentally to achieve
good accuracy on every service -- a key reason for our Bayesian enhancement
in \S\ref{s:bayesian}.

\heading{Correctness Proof Sketch.}
The proof shows that if there were targeting, every non-targeting
input would have a vanishingly small probability to be in a significant fraction of the
active accounts.
Let us call {\em S} the set of inputs contained in a significant fraction of
the active accounts. Without targeting, these inputs would be present
in the accounts by mere chance.  Since inputs are independently distributed into the accounts, we show that the probability of {\em S} not being empty decreases
exponentially with the number of active accounts (through Chernoff bounds).
With targeting, we show that
with high probability no other input than the explaining combination is in {\em S}, because of the bounded noise hypothesis.
Appendix~\ref{s:correctnessproof} provides further proof details.

The proofs and algorithm included in this paper work only for conjunctive combinations (e.g., $D_1$ and $D_2$, see \S\ref{s:formalized_model}).
The theory, however, can be extended to disjunctive combinations
(e.g., ($D_1$ and $D_2$) or $D_5$), but the algorithm for detecting such
combinations is more complex and relies on a recursive argument: if we find one combination from the disjunction,
then the active accounts that include this combination define a context where
the combination appears non-targeting because it is everywhere. 
\ifnum\isTR=1
If we recursively
apply our algorithm in this context, we can detect the second combination in the
disjunction, then the third, etc (see Appendix).
\else
If we recursively
apply our algorithm in this context, we can detect the second combination in the
disjunction, then the third, etc (see technical report~\cite{xray-tr}).
\fi

\subsubsection{Self-Tuning Bayesian Algorithm}
\vspace{-3pt}
\label{s:bayesian}

\begin{figure}
\begin{minipage}{.48\linewidth}
\begin{lstlisting}[mathescape]
// `\textbf{Bayesian Prediction Alg:}`
// Runs with each collected ad.
`\textbf{In}`: Output $O_k$ (e.g. an ad).
`\textbf{Out}`: Targeted input.
// Compute probabilities.
`{\bf foreach}` input $D_i$ do
 $\probaof{D_i \cond \vec{x}}$ = bayes($\probaof{\vec{x} \cond D_i}$)
end
// Compute untargeted prob.
$\probaof{D_\emptyset \cond \vec{x}}$ = bayes($\probaof{\vec{x} \cond D_\emptyset}$)
// Return event with max prob.
return $D_i$ with max $\probaof{D_i \cond \vec{x}}$
\end{lstlisting}
\end{minipage}\hfill
\begin{minipage}{.46\linewidth}
\begin{lstlisting}[mathescape]
// `\textbf{Parameter Learning Alg:}`
// Runs periodically.
// Initialize params (arbitrary).
$p_{in}=.7$,$p_{out}=.01$,$p_{\emptyset}=.1$
do
 `{\bf foreach}` output $O_k$ do
  Run Bayesian Prediction.
 end
 Update $p_{in}$, $p_{out}$, $p_{\emptyset}$
  from predictions.
until $p_{in}$, $p_{out}$, $p_{\emptyset}$ converge
end
\end{lstlisting}
\end{minipage}
\vspace{-0.3cm}
\caption{\small {\bf Bayesian Correlation.}
Left: Bayesian prediction algorithm for behavioral targeting.
Right: typical iterative inference process to learn parameters.}
\label{f:algo-bayes}
\end{figure}

The Set Intersection algorithm provides a good theoretical
foundation; however, it requires parameters be tuned and applies only to
behavioral targeting, not contextual targeting.
Thus, we include in \xray a more robust, self-tuning version
that leverages a Bayesian algorithm to adjust parameters automatically through
iterated inference.  Our algorithm relies on three models: one that predicts
behavioral targeting, one that predicts contextual targeting, and one that combines
the two.

\heading{Behavioral Targeting.}
The Bayesian behavioral targeting model uses the same random assignment as the
Set Intersection algorithm, and it leverages the same information from the shadow
account observations, $\vec{x}$.
It counts the observations $x_j$ of ad $O_k$ in an account $j$ as a binary
signal: if the ad has appeared at least once in account $j$, we count it once;
otherwise we do not count it.
% Both values are considered useful information in the algorithm, an aspect that
% was not true in the Set Intersection algorithm.
Briefly, the Bayesian model is a simple generative model that simulates the audited
service given some targeting associations (e.g., $D_i$ triggers $O_k$).
It computes the probability for this model to generate the outputs we do observe for
every targeting association.  The most likely association will be the one \xray returns.

In more detail if the ad were targeted towards $D_i$, then an account $j$ containing $D_i$ would see this ad at least once with a \emph{coverage} probability $\pin$; otherwise, it would miss it with probability $(1-\pin)$. An account $j'$ without input $D_i$ would see the ad with a smaller
probability, $\pout$, missing it with probability $(1-\pout)$. If the ad were not
behaviorally targeted, it would appear in each account with the same probability,
$\punt$. If we define $\aacct$ as the set of active accounts that have seen the ad,
and $A_i$ as the set of accounts that contain email $D_i$, then we have the following
definitions for the probabilities:
\vspace{-5pt}
\[
\begin{disarray}[c]{rl}
	\probaof{\vec{x}\cond D_i}
	= &
	 \left( \pin \right)^{|A_i\cap A_k|}
	 \left(1- \pin \right)^{|A_i \cap \bar{A_k}|} \\ &
	\times
	\left( \pout \right)^{|\bar{A_i}\cap A_k|}
	\left(1- \pout \right)^{|\bar{A_i}\cap \bar{A_k}|}
	\vf \\

\probaof{\vec{x}\cond D_\emptyset} = &
\left( \punt \right)^{|A_k|}
\left(1- \punt \right)^{|\bar{A_k}|}
\vf
\end{disarray}
\]
where $D_\emptyset$ designates the untargeted prediction.

The preceding formula has an interesting interpretation that is visible if
placed in the equivalent form:
\vspace{-5pt}
\[
\begin{disarray}[c]{rl}
	\probaof{\vec{x}\cond D_i}
	= &
	\left( \pin\right)^{|A_k|} \left( 1-\pout \right)^{|\bar{A_k}|}  \\ & \times
	\left(\frac{1-\pin}{1-\pout}\right)
	 ^{|A_i \cap \bar{A_k}|}
	 \left(\frac{\pout}{\pin} \right)
	^{|\bar{A_i}\cap A_k|}
\end{disarray}
\]
From the point of view of the event $D_i$,
an account found in $A_i \cap \bar{A_k}$ is a false positive (an
ad was expected but was not shown). This should lower the probability,
especially when the \emph{coverage} $\pin$ is close to $1$.
Inversely, an account found in $\bar{A_i}\cap A_k$ acts as a false negative (we
observed an ad where we did not expect it), which should decrease the probability,
especially when $\pout$ is close to $0$.

These formulas let us infer the likelihood of event $D_i$
according to Bayes' rule: \(
\probaof{A \cond B}=\frac{\probaof{B \cond A} \times \probaof{A}}{\probaof{B}}\).
Figure~\ref{f:algo-bayes} shows two algorithms.  First,  the
prediction algorithm (left) predicts the targeting of $O_k$ by computing
the probabilities defined above, applying Bayes' rule, and returning
the input with the maximum
probability. Second, the parameter learning algorithm (right)
computes the variables that those probabilities depend upon ($\pin$, $\pout$,
and $\punt$) using an iterative process. It repeatedly
runs the prediction algorithm for all outputs and re-computes $\pin$, $\pout$,
and $\punt$ based on the predictions.
It stops when the variables converge (i.e., their variation from one iteration
to another is small).

\heading{Contextual Targeting.}
Contextual targeting is more straightforward since it uses content
shown next to the ad.
\xray also uses Bayesian inference and defines the observations as how many
times ad $O_k$ is seen next to email $D_i$.
Our causal model assumes imperfect coverage: if
this ad were contextually targeted towards $D_i$, it would occur next to that email
with probability $\pin<1$ and next to any other email with probability $\pout$.
Alternatively, if the ad were untargeted, our model predicts it would be shown
next to any email with probability $\punt$. Hence, \(
\probaof{\vec{x}\cond\!\! D_i}  = \left( \pin \right)^{x_i} \left( \pout \right)^{\sum_{i'\neq i} x_i'},
\probaof{\vec{x}\cond\!\! D_\emptyset}  = \left( \punt \right)^{\sum_{i} x_i}
\).
For this model, parameters are also automatically computed by iterated inference.

\heading{Composite Model (\xray).}
The contextual and behavioral mechanisms were designed to detect
different types of targeting.
To detect both types, \xray must combine the two scores.
We experimented with multiple combination functions, including a decision tree and the arithmetic average,
and concluded that the arithmetic average yields sufficiently good results.
\xray thus defines the {\em composite model} that averages scores from individual models,
and we demonstrate in \S\ref{s:eval:accuracy} that doing so yields higher recall for no loss in precision.

% Them into a {\em composite model}, which averages scores from individual models.
% Our experimental results, described in \S\ref{s:eval:accuracy}, demonstrate
% that doing so yields higher recall for no loss in precision.

% \input{archi/correlation}

% \input{archi/placement}

\subsection{Input Matching and Placement}
\vspace{-3pt}
\label{s:input_matching}

Our design of differential correlation, along with our logarithmic
results for random input placement, relies on the fundamental assumption that
the probability of getting an ad $O_{1}$ targeted at an input $D_{1}$ in a
shadow account that lacks $D_{1}$ is vanishingly small.  However, when inputs attract the same ads (a.k.a., overlapping inputs), a naive input placement
can contradict this assumption.  Imagine a Gmail account with multiple
emails related to a Caribbean trip. If placement includes Caribbean
emails in every available shadow account, related ads will appear in
groups of accounts with no email object in common.
\xray will thus classify them as untargeted.  Our Amazon experiments showed
\xray's recall dropping from 97\% to 30\% with overlapping inputs (\S\ref{s:eval:input_matching}).

To address this problem, \xray's Input Matching module identifies
similar inputs and directs the Placement Module to co-locate them
in the same shadow accounts.  The key challenge is to identify
similar inputs.  One method is to use content analysis (\eg keywords matching), but this has limitations. First, it is not service agnostic;
one needs to reverse engineer complex and ever-changing
matching schemes. Second, it is hard to apply to
non-textual media, such as YouTube videos.

In \xray, we opt for a more robust, systems technique rooted in the key insight that
we can deduce similar inputs from contextual targeting.  Intuitively,
inputs that trigger similar targeting from the Web service should attract similar
outputs in their context.
The Input Matching module builds and compare inputs' {\em contextual signatures}.
Contextual signature similarity is the distance between inputs (\eg email)
in a Euclidean space, where each output (\eg ad) is a dimension. The coordinate of an
email in this dimension is the number of times the ad was seen in the
context of the email.
\xray then forwards close inputs to the same shadow accounts. Once the placement is
done, behavioral targeting against that email's group can be inferred
effectively.

This input matching mechanism differs fundamentally
from any content analysis technique, such as keyword matching,
because it groups inputs {\em the same way the Web service
does}.\footnote{We call this method ``monkey see, monkey do''
because we watch how the service groups inputs and group them similarly.}
It is robust and very general: we used it on both Gmail and
Amazon without changing a single line of code to change.

\vspace{-8pt}
\section{\xray-based Tools}
\vspace{-8pt}
\label{s:prototype}

To evaluate \xray's extensibility, we instantiated it on Gmail, YouTube, and Amazon.
The engine, about 3,000 lines of Ruby, was first developed for Gmail.
We then extended it to YouTube and Amazon, without any changes to its correlation algorithms.
We did need to do minor code re-structuring, but the experience
felt turn key when integrating a new service into the correlation machinery.

Building the full toolset required non-trivial coding effort, however.
% \xray's correlation engine, about 4,000 lines of code, is written in
% Ruby and runs as a Web service on the Rails platform and with MongoDB as the
% correlation database.  It can be hosted either on the user's machine or as a
% trusted cloud service.
Instantiating \xray for a specific Web service is a three-step process.  First,
the developer instantiates appropriate data models (less than 20 code lines for our
prototypes).  Second, she implements a service-specific shadow account
manager and plugin; care must be taken not be too aggressive
to avoid adversarial service reactions.  While these implementations are
conceptually simple, they require some coding; our Amazon and YouTube
account managers were built by two
graduate students new to the project, and have around 500 lines of
code.  Third, the developer creates a few shadow accounts for the audited
service and runs a small exploratory experiment to determine the
service's coverage.  \xray uses the coverage to estimate the number of
shadow accounts needed for a given input size. All other parameters are
self-tuned at runtime.

\vspace{-8pt}
\section{Evaluation}
\vspace{-8pt}
\label{s:eval}

We evaluated \xray with experiments on Gmail, Amazon, and YouTube.
While Amazon and YouTube provide ground truth for their targeting, Gmail does not.
We therefore manually labeled ads on Gmail and measured \xray's accuracy,
as described in \S\ref{s:eval:workloads} and validated in \S\ref{s:eval:sanity-check}.
We sought answers to four questions:

\begin{enumerate}
\item[{\em Q1}]{\em How accurate are \xray's inference models?} (\S\ref{s:eval:accuracy})
\item[{\em Q2}]{\em How does \xray scale with input size?} (\S\ref{s:eval:scalability})
\item[{\em Q3}]{\em Does input matching reduce overlap?} (\S\ref{s:eval:input_matching})
\item[{\em Q4}]{\em How useful is \xray in practice?} (\S\ref{s:eval:experience})
\end{enumerate}

% After describing our experimental methodology, we address each question in turn.

\begin{figure}[t]
  \centering
  \includegraphics[width=1.0\linewidth]{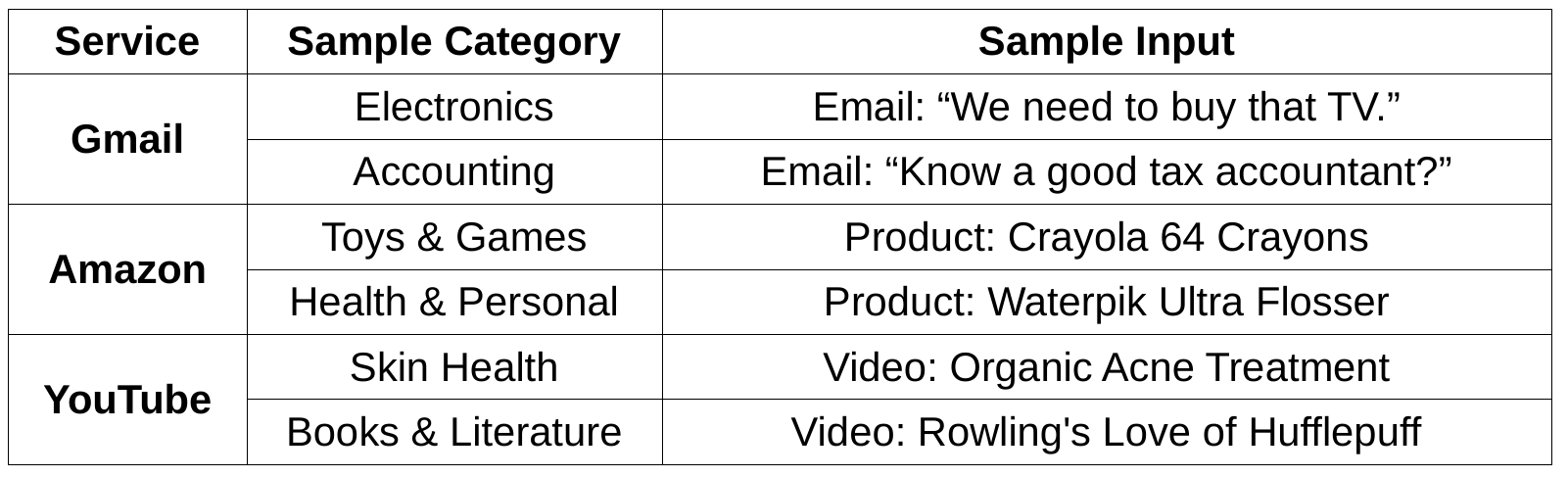}
  \caption{\small {\bf Sample Inputs and Categories.} In total, we developed
  inputs in 64 categories for Amazon and YouTube and in 51 categories for Gmail.}
  \vspace{-0.2cm}
  \label{f:workload}
\end{figure}

\subsection{Methodology}
\vspace{-3pt}
\label{s:eval:workloads}

We evaluated \xray with experiments on Gmail, Amazon, and YouTube.
For inputs,
we created a workload for each service by selecting topics from well-defined
categories relevant for that service.
For Gmail and YouTube, we crafted emails and selected videos based on AdSense categories~\cite{adsense_categ}; for Amazon, we selected products from its own product categories~\cite{amazon_categ}.
\F\ref{f:workload} shows several sample categories and sample inputs in each.
We used these categories for most of our experiments (\S\ref{s:eval:accuracy}--\S\ref{s:eval:input_matching}).
We used these categories to create two types of workloads: (1) a non-overlapping workload, in which
each data item belonged to a distinct category, and (2) an overlapping
workload, with multiple data items per category (described in
\S\ref{s:eval:input_matching}).
% All data collection was performed from the
% same city, and on Gmail we loaded the user account 100 times to obtain a
% reasonable number of ads for each experiment.

To assess \xray's accuracy, we needed the ground truth for associations.  Amazon
and YouTube provide it for their recommendations.
For instance, Amazon provides a link ``Why recommended?'' for each recommendation;
when clicked, it shows an explanation of the form ``The [Coloring Book] is recommended because your wish list includes [Crayola Crayons Set].''
% YouTube explains its video recommendations similarly.
For Gmail, we manually labeled ads based on our
personal assessment.  The ads for different experiments were labeled by
different people, generally project members.
A non-computer scientist labeled the largest experiment (51 emails).
%For our largest Gmail experiment (51 emails), we
%co-opted an independent party (professional, non-computer scientist) to do the
%labeling.
% Because \xray often discovers associations that we as users might never
% think of, our Gmail experiments involved a re-labeling phase where we looked at
% \xray's output to correct any mis-labeling.  We were {\em extremely
% conservative} in acknowledging \xray's correctness over our initial intuition.
% For example, an ad about a hybrid car was first labeled as targeted by the
% {\em New car} email; on closer examination, we observed that the ad appeared
% in XXX accounts whose only common data was our biking-related email.  While
% this re-labeling step changed the results very little (at most 10\%), it
% allowed us to discover interesting targeting as described in
% \S\ref{s:eval:experience}.

% First, we use orthogonal topics so that inputs fall into separate categories and
% do not trigger overlapping targeting. We then relax this assumption in
% \S\ref{s:eval:input_matching} and evaluate \xray on a overlapping workload.

We evaluate two metrics: (1) {\em recall}, the fraction
of positive associations labeled as such, and (2) {\em precision}, the fraction
of correct associations.  We define {\em high accuracy} as having both high
recall and high precision.

\begin{figure}[t]
\centering
{\scriptsize
\tabcolsep=0.12cm
\begin{tabular}{|l|l|c|c|c|}
\hline
{\bf Ad} & {\bf Targeted} & {\bf Detected} & {\bf \xray} & {\bf \# Accounts} \\
{\bf Keyword} & {\bf Email} & {\bf by \xray?} & {\bf Scores} & {\bf \& Displays} \\
\hline
Chaldean  & Like Chaldean  & Yes & 0.99, & 13/13,  \\
Poetry    & Poetry?        &  & 1.0  &  1588/1622   \\ \cline{2-4}
\hline
Steampunk & Fan of Steampunk? & Yes & 0.99,  & 13/13,  \\
          &                   &  & 1.0 & 888/912     \\ \cline{2-4}
\hline
Cosplay   & Discover Cosplay. & Yes & 0.99,  & 13/13,  \\
          &                   &  & 1.0   & 440/442  \\ \cline{2-4}
\hline
Falconry   & Learn about Falconry. & Yes & 0.99,  & 13/13,   \\
           &                  &  & 1.0   & 1569/1608 \\ \cline{2-4}
\hline
\end{tabular}
}
\caption{\small {\bf Self-Targeted Ads.} Fourth column shows \xray's correlation scores
X, Y, the (Bayesian) Behvioral and Contextual scores, respectively.
Fifth column shows raw behavioral and contextual data for better interpretation:
X/Y, Z/T means that the ad was seen in X active accounts that contain the targeted email out of a total of Y active accounts; the ad was shown Z times in the context of the targeted email out of a total of T times.}
\label{f:self-target}
\vspace{-0.2cm}
\end{figure}

\subsection{Sanity-Check Experiment}
\vspace{-3pt}
\label{s:eval:sanity-check}

% Because Gmail, which much of our evaluation is based upon, does not provide
% ground truth, we begin with a small, sanity-check experience 

To build intuition into \xray's functioning, we ran n simple sanity-check experiment on Gmail.
Recall that, unlike Amazon and YouTube, Gmail does not provide any ground truth, requiring us to manually label associations, a process that can be itself faulty.
Before measuring \xray's accuracy against labeled associations, we checked that \xray can detect associations for our own ads, whose targeting we control.
For this, we strayed away from the aforementioned methodology to create a highly controlled experiment.
We posted four Google AdWords campaigns targeted on very specific keywords (Chaldean Poetry, Steampunk, Cosplay, and Falconry), crafted an inbox that included one email per keyword, and used \xray to recover the associations between our ads and those emails.
In total, we saw our ads 1622, 912, 442, and 1608 times, respectively, across all accounts (shadows and master).
\F\ref{f:self-target} shows our results.
After one round of ad collection (which involved 50 refreshes per email), \xray correctly associated all four ads with the targeted email.
It did so with very high confidence: composite model scores were 0.99 in all cases, with very high scores for both contextual and behavioral models.
The figure also shows some of the raw contextual/behavioral data, which provides intuition into \xray's perfect precision and recall in this controlled experiment.
We next turn to evaluating \xray in less controlled environments, for which we
use the workloads and labeling methodology described in \S\ref{s:eval:workloads}.

\begin{figure}[t]
\subfigure[{\bf Recall}]{
  \includegraphics[width=0.47\linewidth]{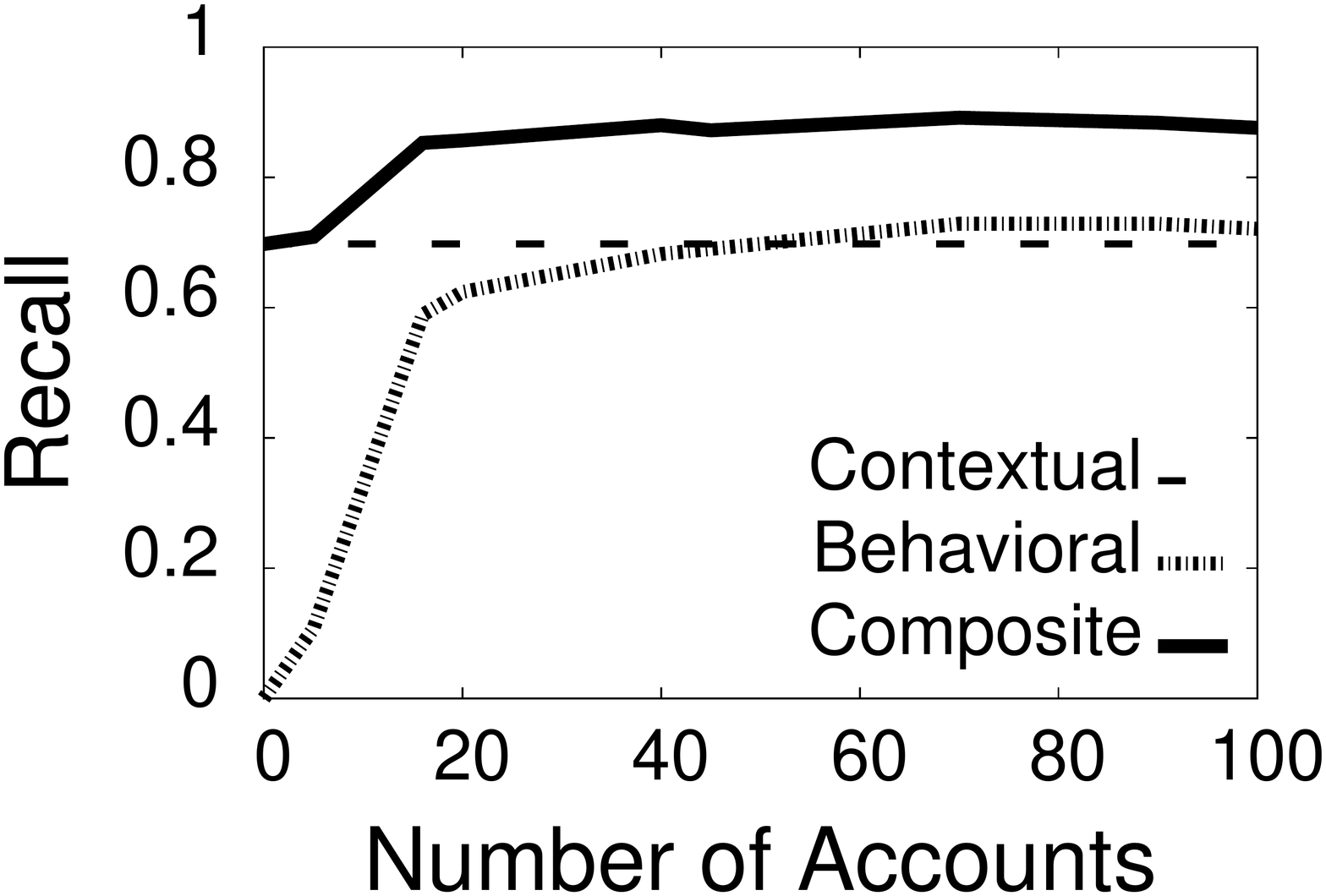}
  \label{f:gmail_recall}
}
\subfigure[{\bf Precision}]{
  \includegraphics[width=0.47\linewidth]{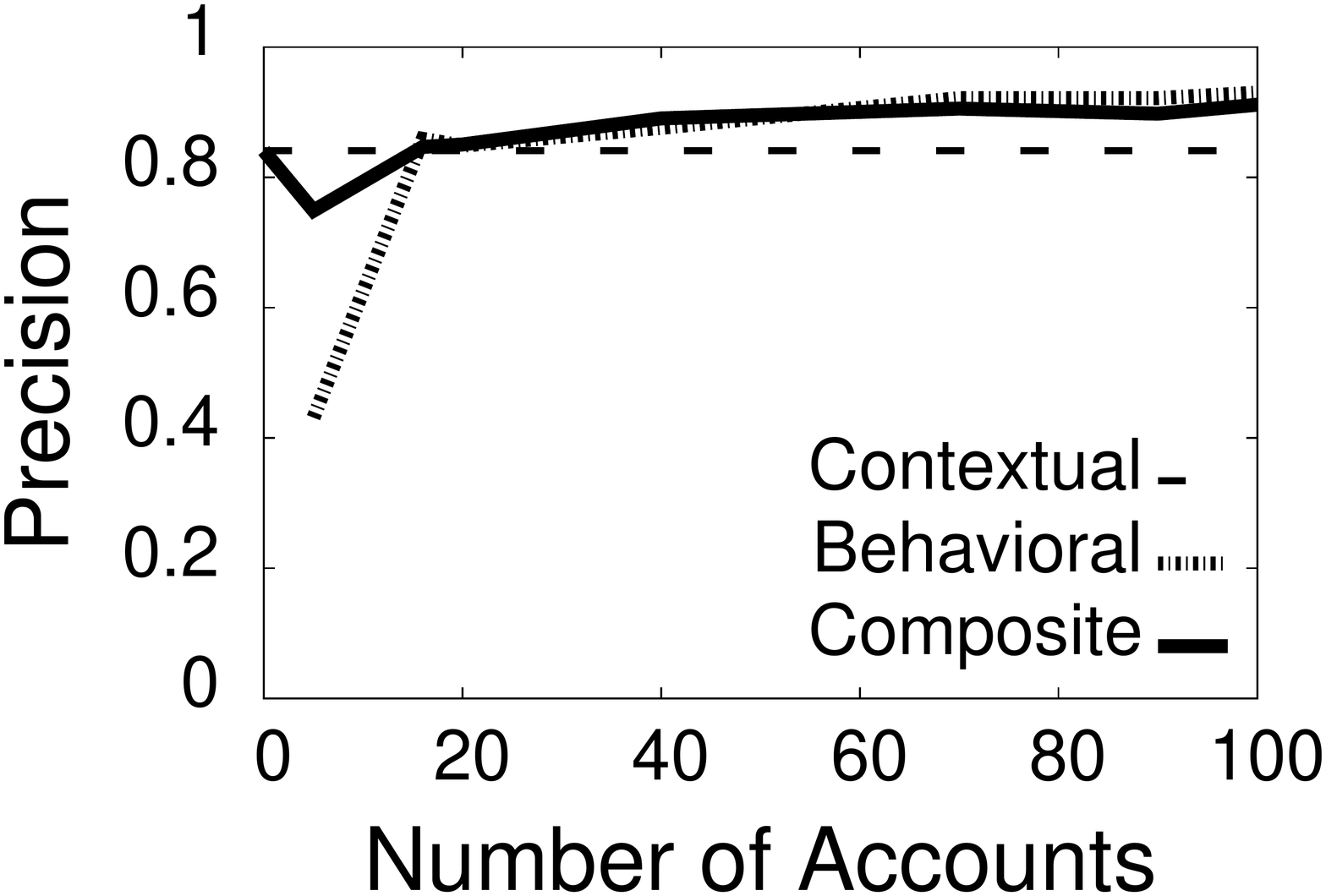}
  \label{f:gmail_precision}
}
\label{f:multifigure}
\vspace{-0.3cm}
\caption{\small {\bf Bayesian Model Accuracy.}
  Recall and precision for each of the three Bayesian models vs. shadow account number, using the Bayesian algorithm. \xray needed 16 accounts to
  reach the ``knee'' with high recall and precision.
}
\end{figure}

\subsection{Accuracy of \xray's Inference Models (Q1)}
\vspace{-3pt}
\label{s:eval:accuracy}

% \heading{\xray's correlation mechanisms.}
To assess the accuracy of \xray's key correlation mechanisms (Bayesian behavioral,
contextual, and composite), we measured their recall and precision under
non-overlapping workloads.
Figures \ref{f:gmail_recall} and ~\ref{f:gmail_precision} show how these two
metrics varied with the number of shadow accounts for a 20-email experiment on
Gmail.   The results indicate two effects.  First, both contextual and
behavioral models were required for high recall.  Of the 193 distinct ads seen
in the user account, 121 (62\%) were targeted, and \xray found 109 (90\%) of
them, a recall we deem high. Of the associations \xray found, 37\% were found
by only one of the models: 15 by the contextual model only, and 24 by the
behavioral model only. Thus, both models were necessary, and composing them
yielded high recall.  Our Amazon and YouTube experiments (which provide ground
truth) yielded very similar results:
on a 20-input experiment, we reached over 90\% recall and precision with only 8
and 12 accounts, respectively.

Second, the composite model's recall exhibited a knee-shaped curve for
increasing shadow account numbers, with a rapid improvement at the beginning
and slow growth thereafter.  With 16 accounts, \xray exceeded 85\%
recall; increasing the number of accounts to 100 yielded a 1.9\% improvement.
Precision also remained high (over 84\%) past 16 accounts.  We
define the {\em knee} as the minimum number of accounts needed to reap most of
the achievable recall and precision.
% The question now becomes how this knee scales with the number of inputs, a topic
% we discuss next.

\begin{figure}[t]
\centering
\subfigure[{\bf Recall}]{
  \includegraphics[width=0.47\linewidth]{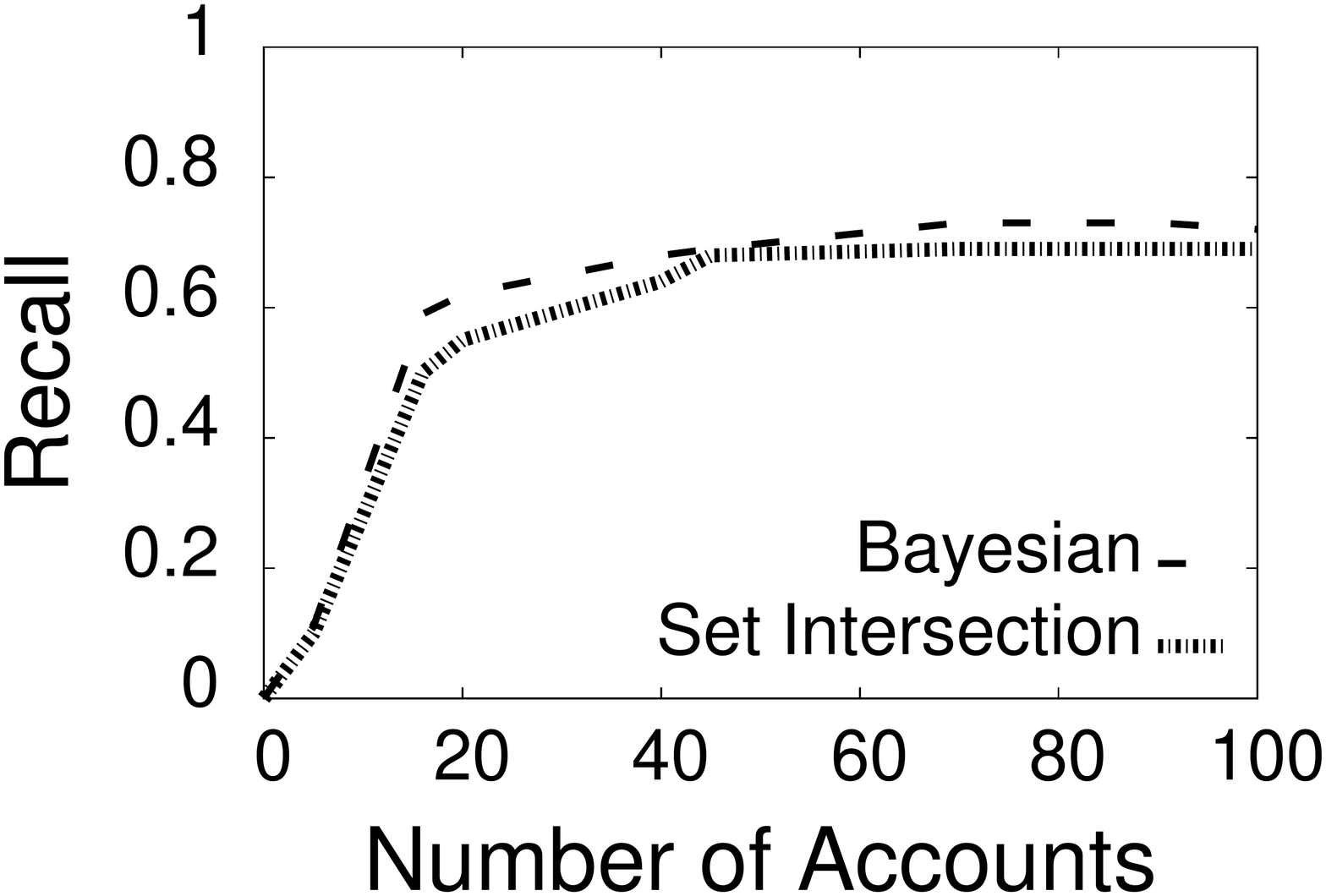}
  \label{f:setbayes_recall}
}
\subfigure[{\bf Precision}]{
  \includegraphics[width=0.47\linewidth]{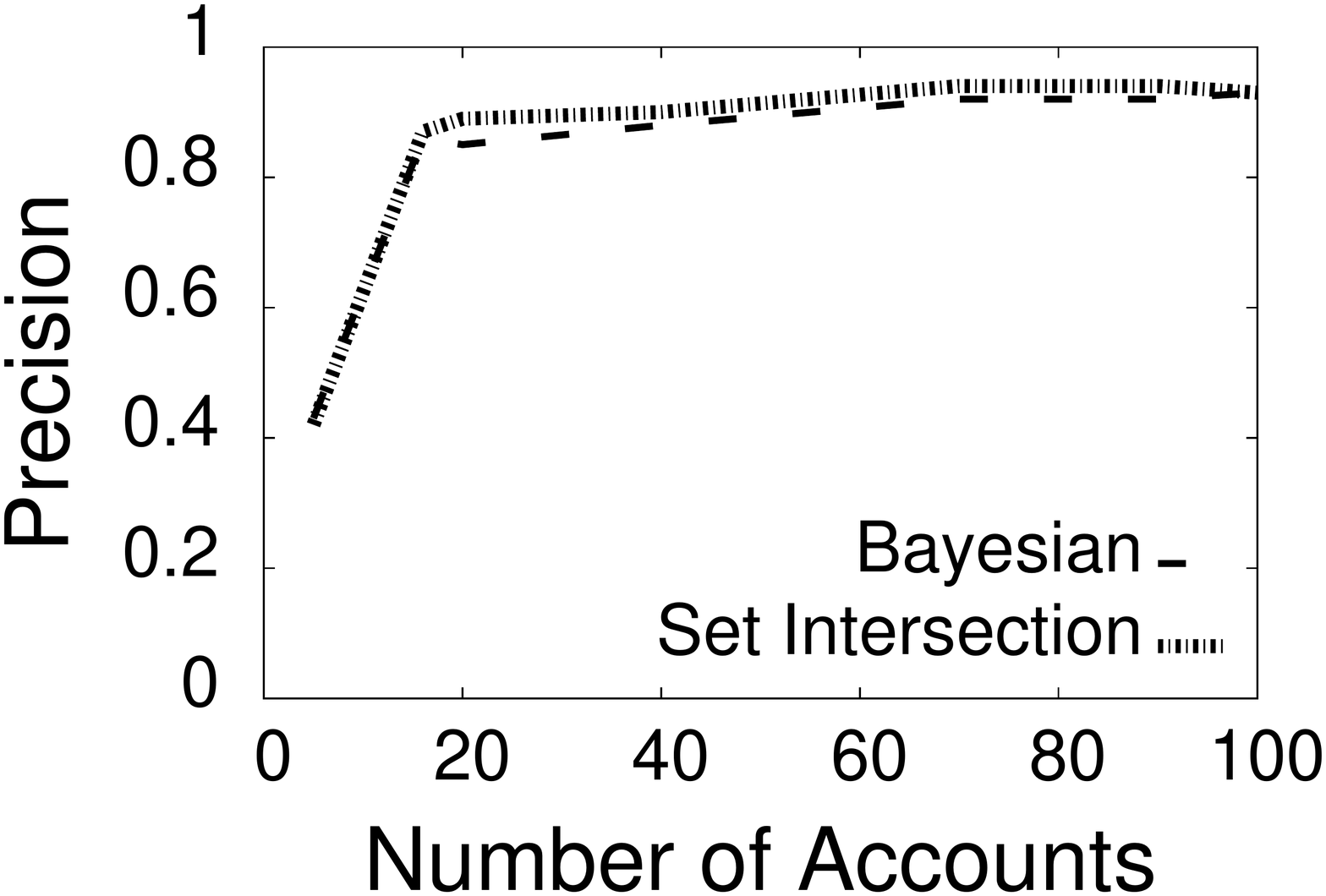}
  \label{f:setbayes_precision}
}
\label{f:multifigure}
\vspace{-0.3cm}
\caption{\small {\bf Bayesian vs. Set Intersection Comparison.}
  Recall and precision for detecting {\em behavioral} targeting with each algo.
}
\end{figure}

% \heading{Bayesian vs. Set Intersection.}
We also wished to compare the accuracy of the Bayesian algorithm,
which conveniently self-tunes its parameters, to the
parameterized Set Intersection algorithm.  We manually tuned the latter
as best as we could.
Figures \ref{f:setbayes_recall} and ~\ref{f:setbayes_precision} show
the recall and precision for detecting behavioral targeting with
the two methods for a non-overlapping workload.  The two algorithms performed similarly,
with the Bayesian staying within 5\% of the manually tuned algorithm.
We also tested the algorithms on an Amazon dataset, and using a version of the Set
Intersection algorithm with empirical optimizations. The conclusion holds: the
Bayesian algorithm, with self-tuned parameters, performs as well as the Set
Intersection technique with manually tuned parameters.
We focus the remainder of this evaluation on the Bayesian algorithm.

\begin{figure*}[t]
\centering
\subfigure[{\bf Scalability with Input Size}]{
  \includegraphics[width=0.3\linewidth]{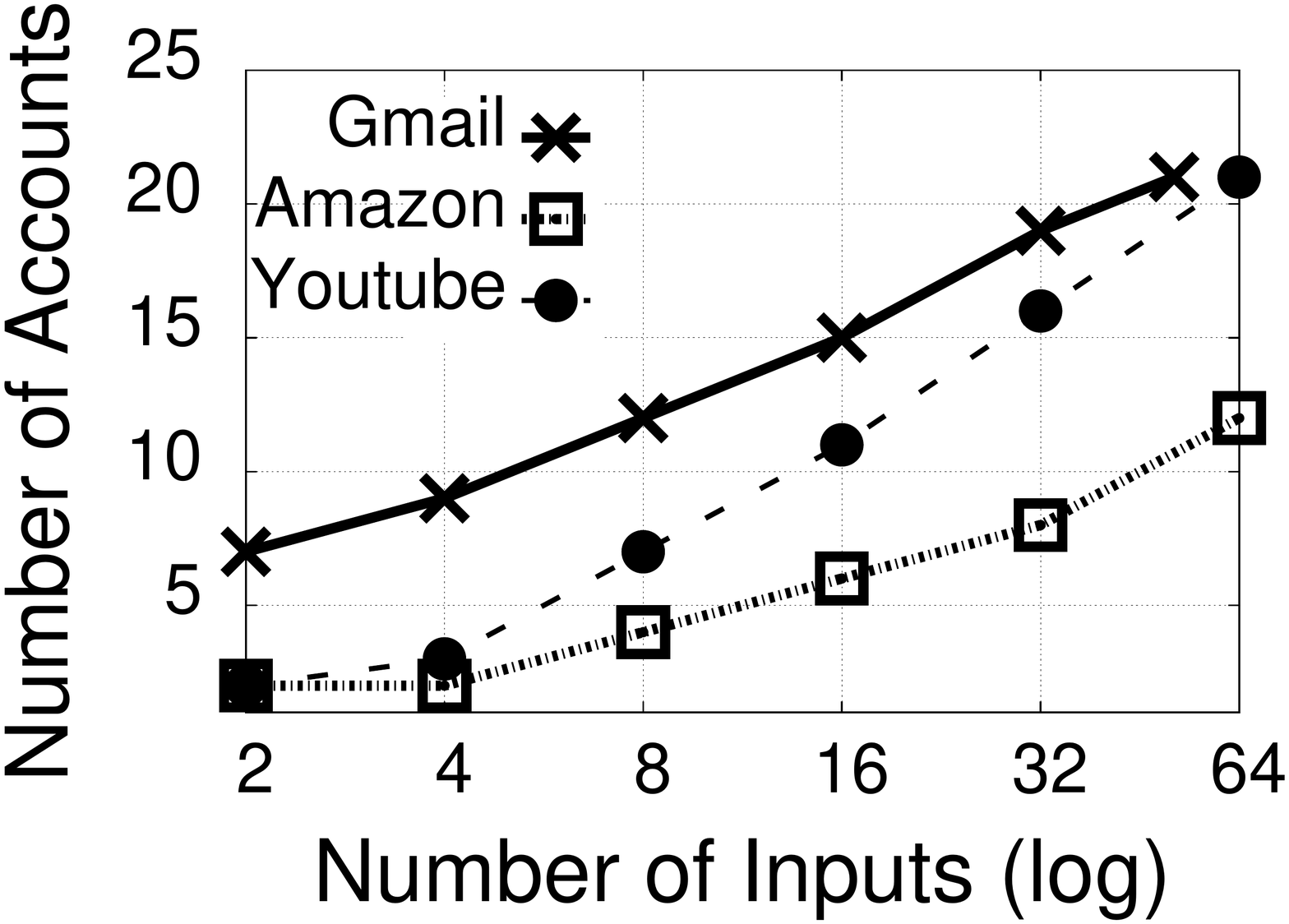}
  \label{f:scaling_gmail_amazon_youtube}
}
\hspace{0.1cm}
\subfigure[{\bf Recall with Input Size}]{
  \includegraphics[width=0.3\linewidth]{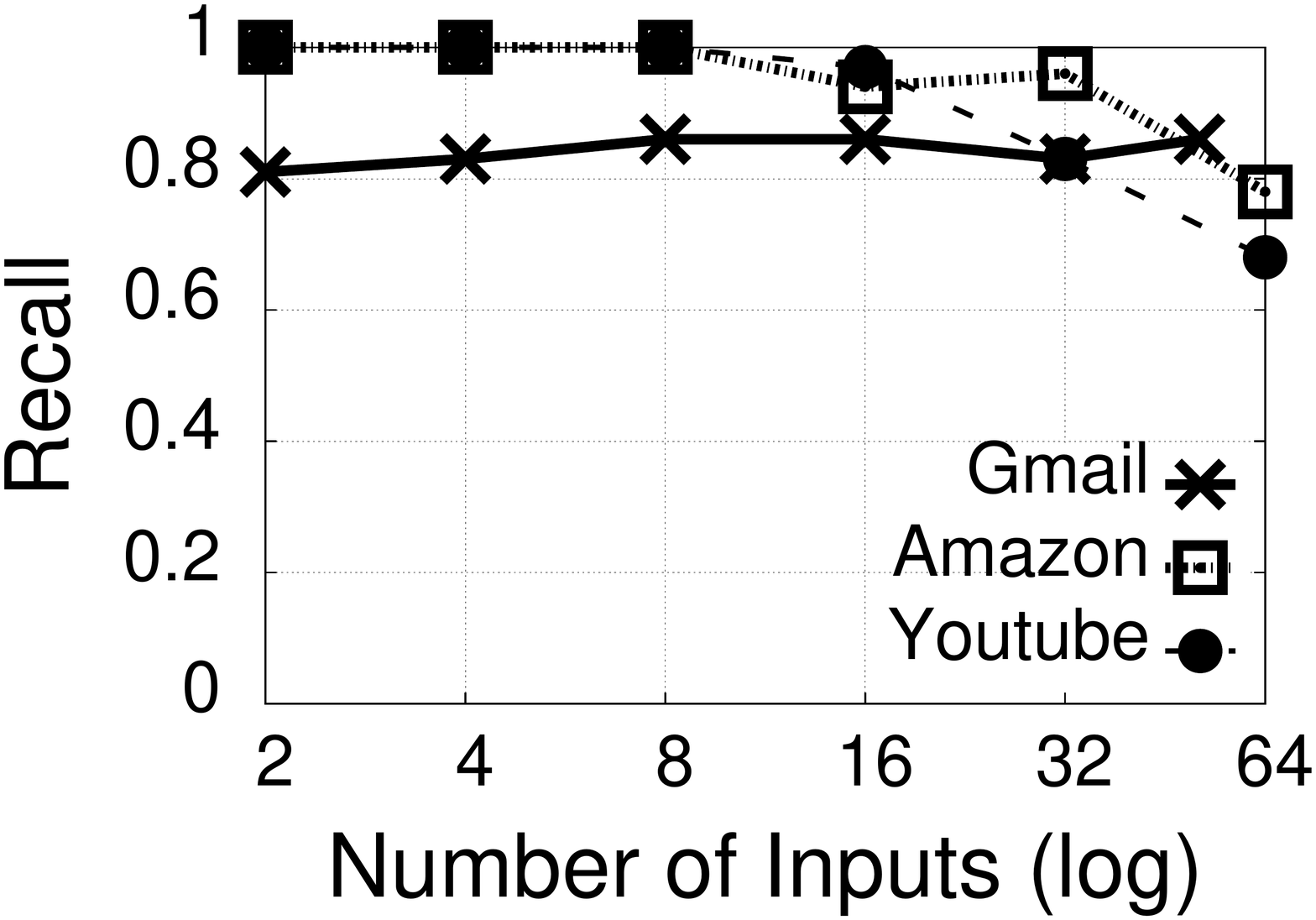}
  \label{f:recall_with_input_gmail_amazon_youtube}
}
\hspace{0.1cm}
\subfigure[{\bf Precision with Input Size}]{
  \includegraphics[width=0.3\linewidth]{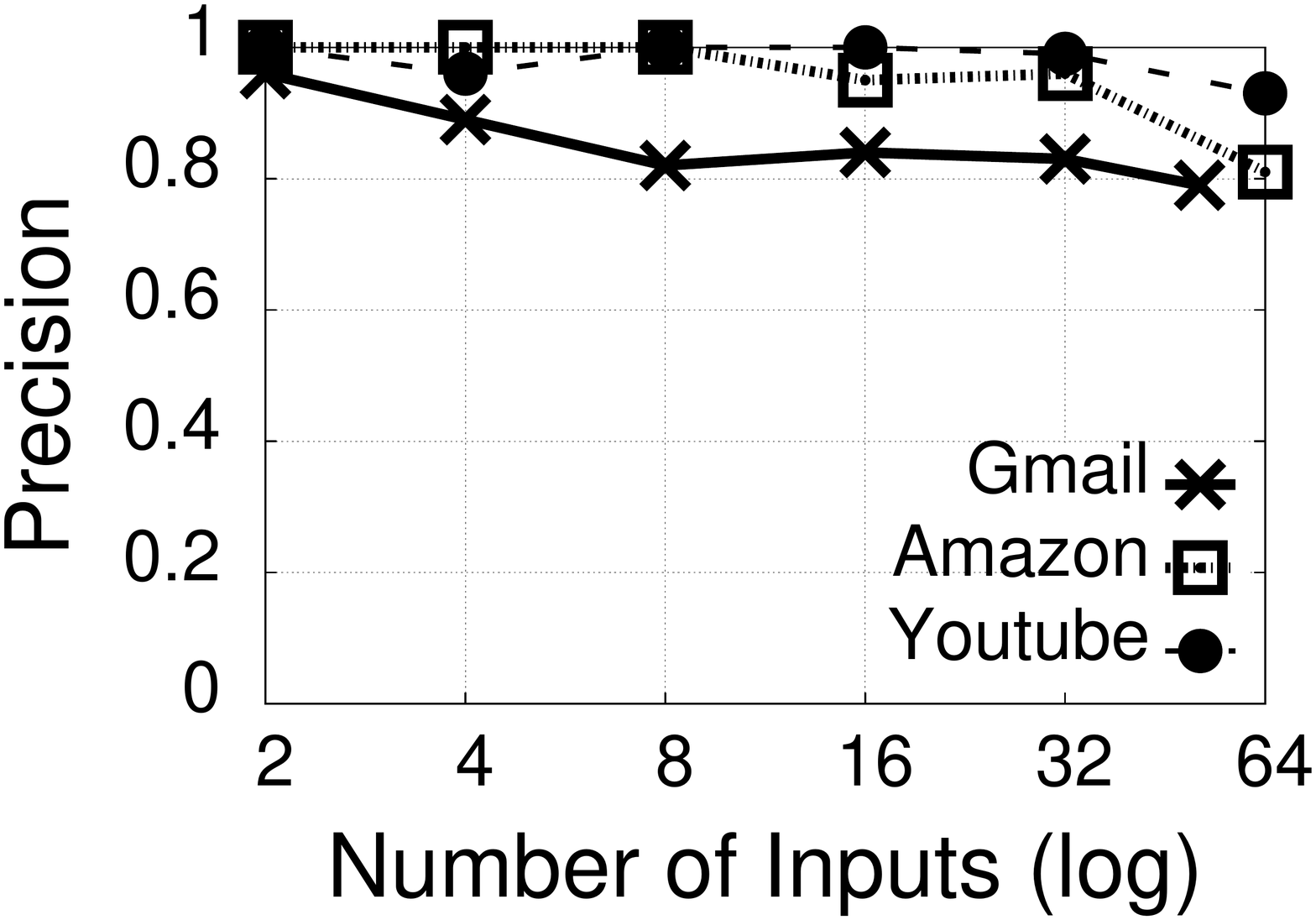}
  \label{f:precision_with_input_gmail_amazon_youtube}
}
%\hspace{0.1cm}
%\subfigure[Coverage]{
%  \includegraphics[width=0.3\linewidth]{results/coverage.pdf}
%  \label{f:coverage_gmail_amazon_youtube}
%}
\vspace{-0.3cm}
\caption{\small {\bf Scalability.}
  (a) Number of accounts required to achieve the knee accuracy for varied
  numbers of inputs.
  (b), (c) Recall/precision achievable with the number of accounts in (a).
  Behavioral uses the Bayesian algorithm.
}
\label{f:scaling}
\end{figure*}

% \begin{figure}[t]
% \centering
% \includegraphics[width=0.7\linewidth]{results/coverage.pdf}
% \caption{\small {\bf Coverage with Input Size.}
% }
% \label{f:coverage_gmail_amazon_youtube}
% \vspace{-0.3cm}
% \end{figure}

\subsection{Scalability of \xray with Input Size (Q2)}
\vspace{-3pt}
\label{s:eval:scalability}
% 4 paragraphs + 3 graphs.

A main contribution of this paper is the realization that, under certain
assumptions, the number of accounts needed to achieve high accuracy for \xray
scales logarithmically with the number of tracked inputs.
We have proven that under certain assumptions, the Set Intersection algorithm scales logarithmically.
This theoretical result is hard to extend to the Bayesian algorithm, so we evaluated it experimentally by
studying three metrics with growing input size: the number of accounts required to
reach the recall knee and the value of recall/precision at this knee.
Figures~\ref{f:scaling_gmail_amazon_youtube},
~\ref{f:recall_with_input_gmail_amazon_youtube}
and~\ref{f:precision_with_input_gmail_amazon_youtube} show the corresponding
results for Gmail, YouTube and Amazon.   For Gmail, the number of accounts necessary to
reach the knee increased less than 3-fold (from 8 to 21) as input size
increased more than 25-fold (from 2 to 51).  For Amazon and YouTube, the increases
in accounts were 6- and 8-fold respectively, for a 32-fold increase in input size.
In general, the roughly linear shapes of the log-x-scale graphs in
\F\ref{f:scaling_gmail_amazon_youtube} confirm the logarithmic increase in the
number of accounts required to handle different inputs.
\F~\ref{f:recall_with_input_gmail_amazon_youtube}
and~\ref{f:precision_with_input_gmail_amazon_youtube} confirm that
the ``knee number'' of accounts achieved high recall and precision (over
80\%).

What accounts for the large gap between the number of accounts needed for high accuracy
in Gmail versus Amazon?  For example, tracking a mere two emails in Gmail
required 8 accounts, while tracking two viewed products in Amazon needed 2
accounts.  The distinction corresponds to the difference in coverage exhibited
by the two services.  In Gmail, a targeted ad was typically seen in a smaller
fraction of the relevant accounts compared to a recommended product in Amazon.
\xray adapted its parameters to lower coverage automatically, but it needed more
accounts to do so.
% The distinction corresponds to the difference in coverage exhibited by the two
% services.  \F\ref{f:coverage_gmail_amazon_youtube} shows how Gmail and Amazon
% coverage (the XX simple definition XX, see \S\ref{s:overview}) varies with the
% number of inputs.  It shows that coverage in Gmail is consistently poorer than
% coverage in Amazon.  \xray adapts to lower coverage automatically (no
% fine-tuning is needed), however it requires more accounts to cope with it.
% As a final note, the figure also provides indication for why \xray's recall and
% precision for Amazon in \F\ref{f:recall_with_input_gmail_amazon_youtube} and
% \F\ref{f:precision_with_input_gmail_amazon_youtube}, respectively drops at 64
% independent items placed in the cart: coverage drops.

Overall, these results confirm that our theoretical scalability results 
hold for real-world systems given carefully crafted, non-overlapping
input workloads. We next investigate how more realistic overlapping input
workloads challenge the accuracy of our theoretical models and how input
matching -- a purely systems technique -- helps address this challenge.

\iffalse
However, the exact values at the knee are not constant.  At the core, the
difference is due to coverage, which decreases with the number of inputs instead
of staying constant as in the theoretical model.
\F\ref{f:coverage_gmail_amazon_youtube} shows how coverage evolves with the
number of inputs.
We can observe two trends. First Gmail's coverage is not as good as
Amazon's, around 0.5 instead of more than 0.8.  This is why Gmail consistently
requires more accounts than Amazon to achieve high accuracy.  Second, Gmail's
coverage oscillates around 0.5, while Amazon's is first very close to 1, and
then starts dropping. This explains the fairly consistent precision and recall
for Gmail, and the observed drop for Amazon.
We expect that this performance degradation will be acceptable for most
scenarios: while user's may have more than 64 items in an Amazon wishlist, it is
unlikely that it will be composed of 64 distinct categories.  We thus need to
match similar inputs by topic to reduce the number of items \xray examines and
improve accuracy, as we show in the next section.
\fi

\begin{figure}[t]
\centering
\subfigure[{\bf Recall}]{
  \includegraphics[width=0.47\linewidth]{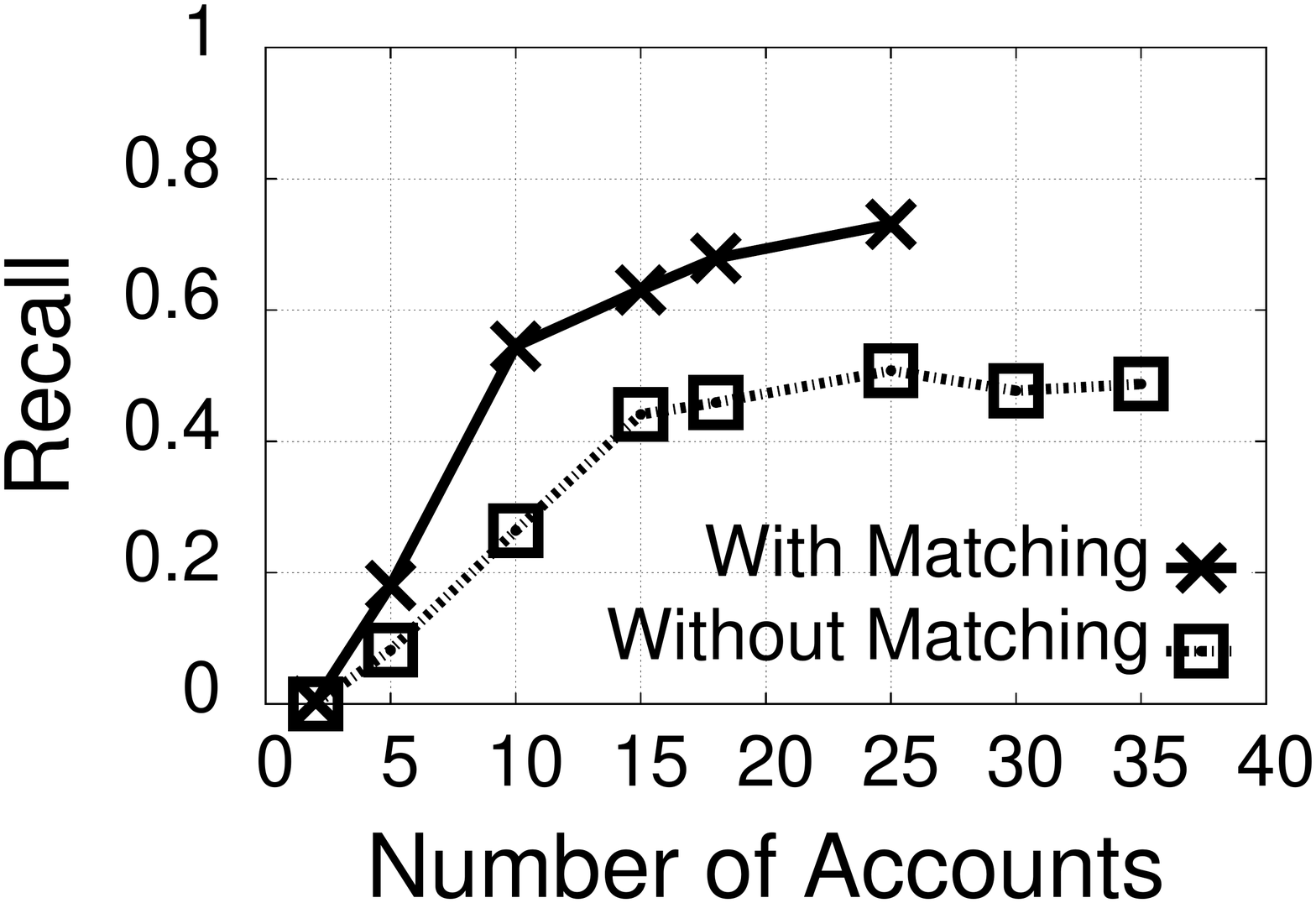}
  \label{f:input_matching_recall_gmail}
}
% \hspace{0.1cm}
\subfigure[{\bf Precision}]{
  \includegraphics[width=0.47\linewidth]{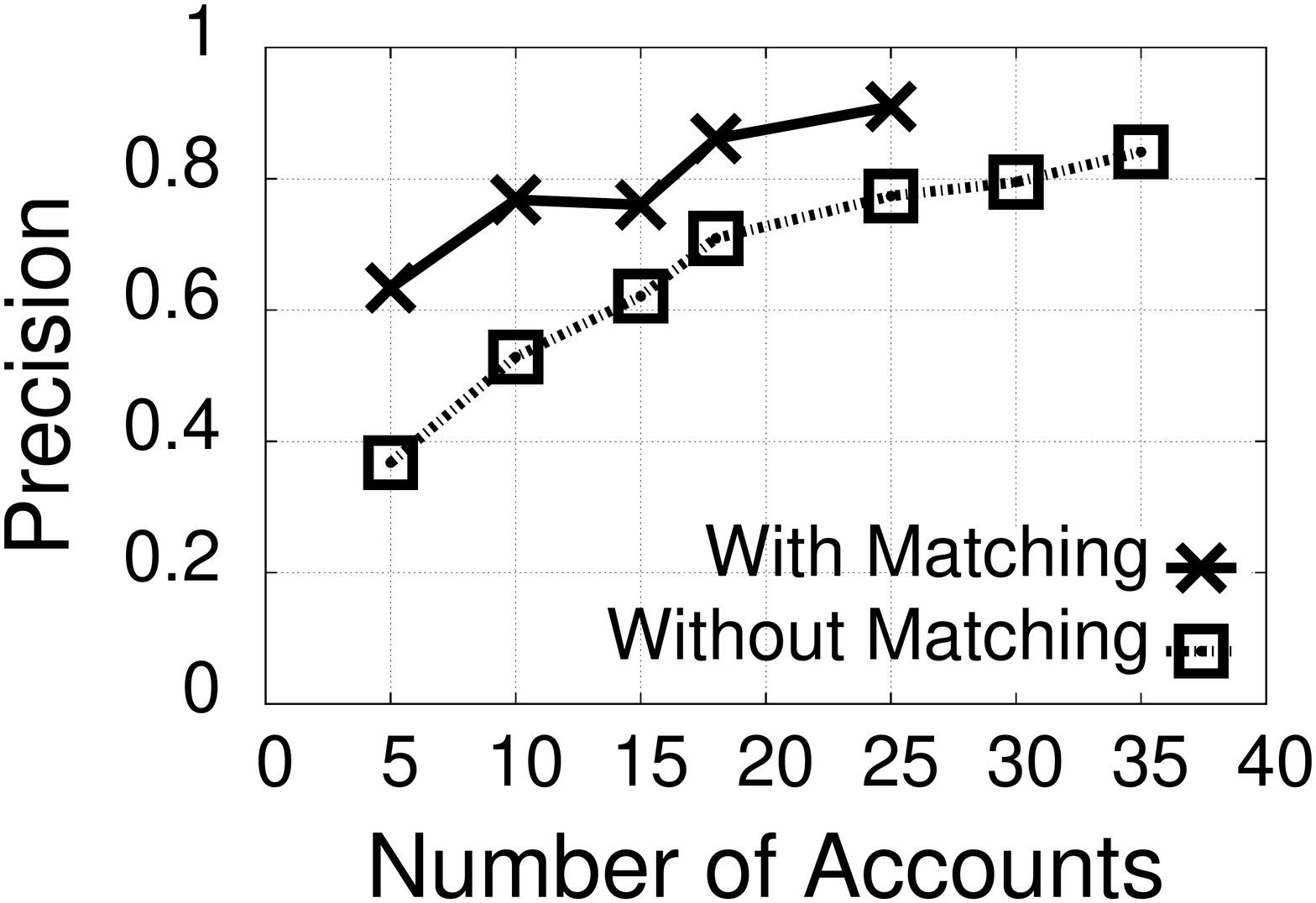}
  \label{f:input_matching_precision_gmail}
}
\vspace{-0.3cm}
\caption{\small {\bf Input Matching effectiveness.}
Behavioral (Bayesian) recall and precision in Gmail with overlapping inputs, with
and without Matching.
}
\label{f:scaling}
% \vspace{-0.3cm}
\end{figure}

\subsection{Input Matching Effectiveness (Q3)}
\vspace{-3pt}
\label{s:eval:input_matching}

To evaluate \xray's accuracy with overlapping inputs, we infused our
workloads with multiple items from the same category (e.g., multiple
emails targeting the same AdSense categories on Gmail and multiple products
in the same category in Amazon).
For the Gmail experiments, we (as users) could not tell when Gmail targeted a specific
email from a group of similar emails.
We therefore ran two different types of experiments:
(1) a controlled, albeit unrealistic, one for Gmail, and (2) a more
realistic one for Amazon.

For Gmail, our controlled experiment replicated various emails {\em
identically} in a user's inbox: 1 email was replicated 4 times, 2 emails 3 times, 4 emails 2 times,
and 12 were single, for a total of 30 emails. This end-of-a-spectrum
workload demonstrates how matching works ideally. \xray matched {\em all}
redundant emails correctly.  More importantly, Figures
\ref{f:input_matching_recall_gmail} and~\ref{f:input_matching_precision_gmail}
show \xray's precision/recall with and without matching-aware placement for
\xray's behavioral model, the only model improved by matching.
Without input matching,  \xray struggled to find differential signals:
even with 35 shadow accounts for a
30-email experiment, recall was only 48\%.  With input matching, \xray's correlation
model drew a stronger signal from each account and attained close to 70\% recall
for 16 accounts.

For Amazon, we created an overlapping workload by selecting three {\em distinct}
products in each of six product categories (e.g., from the
Outdoor \& Cycling category, we selected a helmet, pedals, and shoes).
With a total workload of 18 products, \xray's input matching matched all but one item (shoes) into its correct group. With the new grouping, \xray's recall improved by a factor of 3 (from 30\% to 93\%) compared to
the no-matching case for 18 products with 10 accounts; precision was 2.6
times higher (from 34\% to 88\%).

These results demonstrate that \xray's matching scheme is both portable across
Web services and essential for high accuracy with overlapping workloads.

\subsection{Anecdotal Use Experience (Q4)}
\vspace{-3pt}
\label{s:eval:experience}

\begin{figure}[t]
\centering
{\scriptsize
\tabcolsep=0.11cm
\begin{tabular}{|l|l|c|c|}
\hline
{\bf Topic}  & {\bf Targeted} & {\bf \xray}  & {\bf \# Accounts} \\
             & {\bf Ads}      & {\bf Scores} & {\bf \& Displays} \\
\hline
           & Black Mold Allergy Symptoms?      & 0.99,  & 9/9,    \\
Alzheimer  & Expert to remove Black Mold.      & 0.05  & 61/198   \\ \cline{2-4}
           & Adult Assisted Living.             & 0.99,  & 8/8,   \\
           & Affordable Assisted Living.        & 0.99  & 12/14   \\
\hline
           & Ford Warriors in Pink.     & 0.96,   & 9/9,    \\
Cancer     & Join The Fight.            & 0.98   & 1022/1106  \\ \cline{2-4}
           & Rosen Method Bodywork for   & 0.98,  & 7/7,   \\
           & physical or emotional pain. & 0.05  & 24/598   \\
\hline
           & Shamanic healing over      & 0.99,  & 16/16,    \\
Depression &  the phone.                 & 0.99  & 117/117   \\ \cline{2-4}
           &  Text Coach - Get the girl & 0.93,  & 7/7,   \\
           &  you want and Desire.       & 0.04  & 31/276   \\
\hline
           & Racial Harassment?       & 0.99,  & 10/10,  \\
African    & Learn your rights now.   & 0.2   & 851/5808 \\ \cline{2-4}
American   & Racial Harassment,       & 0.99,  & 10/10,    \\
           & Hearing racial slurs?    & 0.2   & 627/7172     \\
\hline
           & SF Gay Pride Hotel.            & 0.99,     & 9/9,  \\
Homosexuality & Luxury Waterfront.          & 0.1      & 50/99     \\ \cline{2-4}
           & Cedars Hotel Loughborough,     & 0.96,     & 8/8,    \\
           & 36 Bedrooms, Restaurant, Bar.  & 1.0      & 36/43    \\ \cline{2-4}
\hline
           & Ralph Lauren Apparel.         & 0.99,      & 10/10,         \\
           & Official Online Store.        & 0.6       & 85/181  \\ \cline{2-4}
Pregnancy  & Clothing Label-USA.           & 0.99,      & 9/9,            \\
           & Best Custom Woven Labels.     & 1.0       & 14/14   \\ \cline{2-4}
           & Find Baby Shower Invitations. & 0.99,     & 9/9,    \\
           & Get Up To (60\% Off) Here!   &  1.0      & 22/22    \\ \cline{2-4}
\hline
           & Law Attorneys specializing       & 0.99, & 9/9,  \\
Divorce    & in special needs kids education. & 0.99 & 635/666 \\ \cline{2-4}
           & Cerbone Law Firm, Helping        & 0.99, & 10/10,    \\
           & Good People Thru Bad Times       & 1.0 & 94/94     \\
\hline
               & Maui Beach Weddings Serving.  & 0.99, & 7/7,  \\
Enough with    & Affordable  ceremonies.       & 0.0  & 2/728  \\ \cline{2-4}
this marriage  & Romantic Wedding.             & 0.99, & 8/8,    \\
               & Ceremony Planning.           & 0.04 & 4/31     \\
\hline
              & Take a New Toyota Test Drive,     & 0.99, & 7/7,  \\
              & Get a \$50 Gift Card On The Spot. & 0.9  & 58/65     \\ \cline{2-4}
Debt          & Great Credit Cards Search.       & 0.99, & 9/9,   \\
              & Apply for VISA, MasterCard...    & 0.0  & 151/2358\\ \cline{2-4}
              & Stop Creditor Harassment,         & 0.99, & 8/8,    \\
              & End the Harassing Calls.         & 0.96 & 256/373     \\
\hline
\end{tabular}
}
\caption{\small {\bf Example Targeted Ads Uncovered by \xray.} Columns three and four
show the same data as columns four and five in \F\ref{f:self-target}.}
\label{f:experience}
\vspace{-0.25cm}
\end{figure}

To gain intuition into \xray's value in practice, we ran a small-scale, anecdotal experiment that looked for ads attracted by a few specific topics in Gmail.
We created emails focused on a few topics, including cancer, Alzheimer, depression, HIV, race, homosexuality, pregnancy, divorce, debt, and others.
Each email consisted of a number of keywords closely related to one topic (e.g., the depression-related email included {\em depression}, {\em depressed}, and {\em sad}; the homosexuality email included {\em gay}, {\em homosexual}, and {\em lesbian}).
We then launched \xray/Gmail's ad collection several times at intervals of two days, and examined its targeting associations.
\F\ref{f:experience} shows example \xray associations for each of the topics we considered, along with its confidence scores and some of the raw data behind its scores (see \S\ref{s:}).
We conservatively show only a select few of the ads we gathered, for which \xray's confidence -- particularly in behavioral score -- was particularly high.
While our experiment is too small to draw definitive and detailed conclusions about ad targeting in Gmail, we make three high-level observations from our experience.

First, our small-scale experiment confirms that it is possible to target sensitive topics in users' inboxes.
For example, all disease-related emails, except for the HIV-related one, correlated very strongly with various ads.
For example, 
Pregnancy, homosexuality, race, divorce, and debt also attracted ads.
Interestingly, our experience suggests that disease- and 
Overall, we have been surprised .
For instance, ads 7/8, 15/16, and 17 target race, sexual orientation and cancer,
respectively.  The ads we observed were mostly benign or even positive.
However, if no keyword in the ad suggested relation with sensitive topics (\eg ad 17), a
users clicking on the ad may not realize that they could be disclosing private
information to advertisers.  This case inspired Scenario 2 in \S\ref{s:scenarios}.
Suppose an insurance company wanted to gain insight
into pre-existing conditions of its customers before signing them up. It could
create two ad campaigns -- one that targets cancer and another youth -- and
assign different URLs to each campaign. It could then offer higher premium
quotes to visitors who come through the cancer-related ads to discourage them
from signing up while offering lower premium quotes to those who come through
youth-related ads.
% While the ad service (e.g., Google) is not malicious
% in this case,  it is a platform for opening users to such privacy violations, and
% it does so opaquely.
% \xray redresses this type of situation.

Second, our experiments suggest that some advertisers use targeting
capabilities to focus their campaigns on vulnerable subgroups.
In one case, a shamanic phone healing service heavily targeted keywords in our
depression email (ad 1).
In another case, our ``broke'' email attracted many personal loan offers (ad 10)
and deals with high scam potential (ad 11).
Whether these practices are fair is beyond the scope of this work,
but we believe that informed users are empowered users.

Third, many cases, targeting did not have a good semantic
understanding of the emails.
For instance, an email about divorce, that also contained the word marriage
received many ads about wedding ceremonies, like ad 13.
The TV Show email also contained the word
``watch'' and hence got targeted heavily by watch brands (ads 5 and 6).
Context does not seem to be used to disambiguate specific keywords.
We could not tell if the targeting algorithm were incapable of such semantic analysis,
or if the feature were not exposed or used by advertisers.

These results show probable correlations, although we cannot be sure that they denote
targeting. However, we selected only those cases with strong evidence of
correlation between email and ad.

\subsection{Summary}
\vspace{-3pt}
\label{s:summary}

Our evaluation results show that \xray supports fine-grained, accurate data
tracking in popular Web services, scales well with the size of data being
tracked, is general and flexible enough to work efficiently for three Web
services, and robustly uses systems techniques to discover associations when ad
contents provide no indication of them.
We next discuss how \xray meets its last goal: robustness against
honest-but-curious attackers.

\vspace{-8pt}
\section{Security Analysis}
\vspace{-8pt}
\label{s:security_analysis}

As stated in \S\ref{s:threat_model}, two threat models
are relevant for \xray and applicable to different use cases.  First, an {\em
honest-but-curious} Web service does not attempt to frustrate \xray, but it could
incorporate defenses against typical Web attacks, such as DDoS or spam, that might
interfere with \xray's functioning.
% Relevant use cases are given in Scenarios 1 and 2 of
% \S\ref{s:scenarios} and the attack described in \S\ref{s:eval:experience}.
Second, a {\em malicious service} takes an adversarial
stand toward \xray, seeking to prevent or otherwise disrupt its correlations.
Our current \xray prototype is robust against the former threat and can be
extended to be so against the latter.  In either case, third-party advertisers are
untrusted and can attempt to frustrate \xray's auditing.  We discuss each
threat in turn.

\heading{Non-Malicious Web Services.}
Many services incorporate protections against specific automated
behaviors.  For example, Google makes it hard to create new accounts, although
doing so remains within reach.
% (e.g., our phone-based verification tool, which
% buys phone numbers from Twilio, created Google accounts for 7 cents each).
Moreover, many services actively try to
identify spammers and click fraud.  Gmail includes
sophisticated spam filtering mechanisms, while YouTube rate limits
video viewing to prevent spam video promotion.  Finally, many services
rate limit access from the same IP address.

\xray-based tools must be aware of these mechanisms and scale back their
activities
to avoid raising red flags.  For example, our \xray-based tools for Gmail,
YouTube, and Amazon rate limit their output collection in the
shadow accounts.  More importantly, \xray's very design is sensitive to these
challenges:  by requiring as few accounts as possible, we minimize: (1) the
load on the service imposed by auditing, and (2) the amount of input replication
across shadow accounts.  Moreover, \xray's workloads are often
atypical of spam workloads.  Our \xray Gmail plugin sends
emails from one to a few other accounts, while spam is sent from one account to
many other accounts.

\heading{Malicious Third-Party Advertisers.} Third-party advertisers
have many ways to obfuscate their targeting from \xray, particularly if it may
arouse a public outcry. First, an advertiser could
purposefully weaken its targeting by, for example, targeting the same ad 50\% on one
topic and 50\% on another topic. This weakens input/output correlation and may
cause \xray to infer untargeting. However, it also makes the advertisers'
targeting less effective and potentially more ambiguous if their goal is to
learn specific sensitive information about users.
Second, an advertiser might target complex combinations of inputs that
\xray's basic design cannot discover.
\ifnum\isTR=1
We show in Appendix an example of how advertisers might use it, and that our theoretical results extend to those combinations. 
\else
Our accompanying technical report shows an example of how advertisers might
achieve this~\cite{xray-tr}. 
\fi
It also extends our theoretical models
so they can detect targeting on linear combinations with only a constant factor
increase in the number of accounts.
We plan to incorporate and evaluate these extensions in a future prototype.
% Third, an advertiser could inspect specific \xray tools and target ads against
% untracked inputs in combination with tracked ones.  To defend, we recommend that
% \xray tools randomly configure for tracking a number of perhaps uninteresting
% inputs.

\heading{Malicious Web Services.} A malicious service could identify and
disable shadow accounts. Identification could be based on abnormal traffic
(successive reloads of email pages), data distribution within accounts (one
account with lots of data, several others with subsets of it), and
perhaps more. %More generally, spam account identification~\cite{Cao:2012nsdi}
% is an active area of research, and could potentially be adapted to detect \xray
% accounts.
\xray could be extended to add randomness and deception (e.g., fake emails
in shadow accounts, vary email copies).   More importantly, a collaborative
approach to auditing, in which users contribute their ads and input topics
in an privacy-preserving way is a promising direction for strengthening robustness
against attacks.  Web services cannot, after all, disable legitimate user
accounts to frustrate auditing.  We plan to pursue this direction in future work.
% They could, however, fork the users' views
% of the targeted outputs, serving different outputs against different users for
% the same targeted content. This would prevent collaborative correlation;
% however, it might also averserly affect revenue streams from advertising.
% Moreover, a meta-auditing system that watches for such separations could detect
% misbehaviors.

\vspace{-8pt}
\section{Discussion}
\vspace{-8pt}
\label{s:discussion}

\xray takes a significant step toward providing data management transparency in Web services.
As an initial effort, it has a number of limitations.
% First, \xray detects correlation not causality.  For instance, we can
% % imagine a case where a user writes an email on a topic in reaction to an ad that was just
% displayed. In this case the causality is opposite to the one we want to detect.
% The current version of our prototype performs measurements and analysis before sending any new
% email: for every analysis, we know that the email precedes the ad.
First, both the Set Intersection and Bayesian algorithms assume independent
targeting across accounts and over time.  In reality, ad targeting is not always
independent across either.  For example, advertisers set daily ad budgets.
When the budget runs out, an ad can stop appearing in accounts
mid-experiment even though it has
the targeted attributes.  The system might incorrectly assume that no
targeting is taking place, when it could resume the next day.
\xray takes reduced coverage into account, but differences between ads can let
some targeting pass unnoticed.
\xray does not currently account for these dependencies, but estimating their impact
is an important goal for future work.

Second, we assume that targeting noise is bounded and smaller than the targeting
signal. While this condition seems to hold on the evaluated services, other services
making more local decisions may be harder to audit. For instance, a social network
(\eg Facebook) could target ads based on friends' information. The noise created
by the environment could potentially be as high as the targeting signal.
A future solution might be to create shadow accounts with the same friends or shadows of friends.

Third, \xray uses Web services atypically.
To the best of our knowledge, it does not violate any terms of service.
It does, however, collect ads paid for by advertisers to detect correlation.
Ad payment is per impression and pay per click.
The former is vastly less expensive than the latter~\cite{olejnik2013selling}.
\xray creates false impressions only but never clicks on ads. A back-of-the-envelope
calculation using impression pricing from \cite{olejnik2013selling} of \$0.6/thousand impressions reveals that \xray's cost should be minimal: at most 50 cents per ad for our largest experiments.
% Even our largest experiments (1.4M impressions
% of 3.2K unique ads) maintain the cost per ad under \$0.5.

% CPM (Cost Per Mille) cost for a thousand impressions ~ \$0.6
% Biggest experiments:
% 1.4M displays -> \$900
% 3.2k ads
% Most Experiments:
% 120k displays -> \$70
% 500 ads
% For all: less than half a dollar per ad

Despite these limitations, \xray has proven itself useful for many
needs, particularly in an auditing context.  An auditor can craft inputs that
avoid many of these limitations.  For example, emails can be written to avoid
as much overlap as possible and keep the size of inputs used for targeting within
reasonable bounds.  We hope that \xray's solid correlation components will streamline
much-needed investigations -- by researchers, journalists, or the FTC -- into
how personal data is being used.

% \vspace{-10pt}
\vspace{-8pt}
\section{Related Work}
\vspace{-8pt}
\label{s:relwork}

\iffalse
As the widespread deployment of advertising networks have transformed the Web, privacy~\cite{Krishnamurthy:2009bn} and regulatory~\cite{Mayer:2012wt} concerns have made is essential to identify (and sometimes to evade) the ubiquitous tracking of our online activity. Perhaps the most recent active research direction in this area is to minimize information leakage to third parties while maintaining the personalization of ads~\cite{Toubiana:2010tm,Saikat:2011tt}, search~\cite{Fredrikson:2011dn,Toubiana:2011uv}, and social Web widgets~\cite{Roesner:2012uj,Kontaxis:2012vd}. \xray aims to protect users by improving their visibility and by enabling independent third parties to keep online services under scrutiny at all times~\cite{Diakopoulos:2014wo}.
\fi

While \S\ref{s:alternative_solutions} covered Web data protection and auditing
related works, we next cover other related topics.  Our work relates to recent
efforts to measure various forms of personalization, such as
search~\cite{Hannak:2013uk,Xing:2014ws}, pricing~\cite{Anonymous:2012wi}, and ad
discrimination~\cite{Sweeney:2013cw}.  These efforts start from the assumption
that personalization has a dark side (\eg~censorship~\cite{Xing:2014ws}, filter
bubble~\cite{Hannak:2013uk}). They generally employ a methodology similar in
spirit to differential correlation, but their goals differ from ours.
They aim to quantify \emph{how much} output is personalized and what \emph{type}
of information is used overall (be it a user's geography, demographic
attributes, or past behavior).  In contrast, \xray seeks to provide fine-grained
diagnosis of which \emph{input data} generates which personalized results.
Through its scaling mechanisms -- unique in the personalization and data
tracking literature -- \xray scales well even when the relevant inputs are many
and unknown in advance.

Our work also relates to a growing body of research measuring advertising
networks. These networks, notably complex and difficult to
crawl~\cite{Barford:ug}, are rendered opaque by the need to combat click
fraud~\cite{Dave:2012bu}, and have been shown to be susceptible to
leakage~\cite{Korolova:2010to} and profile reconstruction
attacks~\cite{Castelluccia:2012vl}.
As for other personalization, prior studies have focused mostly on macroscopic trends
(\eg What fraction of ads are targeted overall?)~\cite{Barford:ug} or
qualitative trends (\eg Which ads are targeted toward gay
males?)~\cite{Guha:2010hk}. Various studies showed traces -- but not a
prevalence -- of potential abuse through concealed targeting~\cite{Guha:2010hk}
and data exchange between services~\cite{Wills:wf}. These works primarily focus
on display advertising, and each distinguishes contextual advertising using a
specific classifier with semantic categories obtained from Google's Ad
Preferences Managers or another public API~\cite{BinLiu:2013jn}.

\xray departs significantly from these works. First, since it entirely
ignores the content and even the domain of targeting, it is readily applied as-is
to ads in Gmail, product recommendations, and videos. Second, while previous
methods label ads as ``behavioral'' in bulk once other explanations
fail~\cite{BinLiu:2013jn}, \xray remains grounded on positive evidence of
targeting, and it determines to \emph{which} inputs an output should be attributed.
Third, \xray's mechanisms to avoid exponential input placement and deal with
overlapping inputs are unprecedented in the Web-data-tracking context. While they
resemble \emph{black box} software testing~\cite{Beizer:1995ty}, the specific
targeting assumption we leverage have, to our knowledge, no prior equivalent.

%Finally, our work relates to an enormous body of work on taint tracking systems~\cite{hails, taintdroid, other_stuff}.  These systems all assume a controlled runtime environment, such as a modified operating system, language, or runtime.  \xray tracks data in the open, uncontrolled Web by relying purely on correlation.  \xxx{taint tracking based on correlation.}

\vspace{-8pt}
\section{Conclusions}
\vspace{-8pt}
\label{s:conclusion}

The tracking of personal data usage poses unique
challenges. \xray shows for the first time that accurate, \emph{fine-grained}
tracking need not compromise portability and scalability. For users
who care about \emph{which} piece of their data has been targeted, it offers a
unique level of precision and protection. Our work calls for and promotes the best
practice of voluntary transparency, while at the same time empowering
investigators and watchdogs with a significant new tool for increased vigilance.

\vspace{-8pt}
\section{Acknowledgements}
\vspace{-8pt}

We extend special thanks to our shepherd, Dan Boneh, for his valuable guidance.
We also thank the anonymous reviewers and numerous colleagues who have given us
feedback, including: Jonathan Bell, Sandra Kaplan, Michael Keller, Yoshi Kohno, Hank Levy, Yang Tang, Nicolas Viennot, and Junfeng Yang.
This work was supported by funds from DARPA Contract FA8650-11-C-7190, NSF CNS-1351089, Google, and Microsoft.
% The views and opinions expressed in this paper belong solely to the authors.

% -------------------- %

\renewcommand{\baselinestretch}{0.9}
{
  \footnotesize
  \bibliographystyle{abbrv}
  \bibliography{abbrev,conferences,refs,news,my_papers,AllPapers}
}

% \cleardoublepage
\appendix
\vspace{-8pt}
\section{Proof of Theorem~\ref{res:placement}}
\vspace{-8pt}
\label{s:correctnessproof}

\newcommand{\comb}{\mathcal{C}}
\newcommand{\combs}{\mathcal{S}}
\newcommand{\rmax}{r_{\textrm{max}}}
\newcommand{\lmax}{l_{\textrm{max}}}
\newcommand{\famad}{{S^{(\textrm{ad})}}}
\newcommand{\famadin}{{S^{(\textrm{ad,in})}}}
\newcommand{\famadout}{{S^{(\textrm{ad,out})}}}
\newcommand{\famcore}{{S^{(\textrm{core})}}}
\newcommand{\famdif}[1]{{\Delta^{(\textrm{ad})}}\left( #1 \right)}
\newcommand{\famdifin}[1]{{\Delta^{(\textrm{ad,in})}}\left( #1 \right)}
\newcommand{\famdifcore}[1]{{\Delta^{(\textrm{core})}}\left( #1 \right)}

\subsection{Targeting functions, Axioms and Core Family}
\vspace{-3pt}

\ifnum\isTR=1
To formalize our main result we need to carefully define how targeting works and the simple qualitative axioms that it obeys. We show in this section that, provided those axioms are satisfied, targeting can always be associated with a small number of input combinations that we call its core.

\subsubsection{Definitions and main result}
\fi

Given a fixed universe of $N$ inputs, a \emph{combination} $\comb$ of order $r$, also called $r$\_combination, is a subset of $r$ elements among the $N$ inputs. 

Each given ad is associated with a \emph{targeting function} defined as a mapping $f$ from any subset $\comb$ of the $N$ inputs into $\{0,1\}$, where $f(\comb)=1$ denotes that an account containing $\comb$ as inputs should be targeted. By convention, untargeted ads are associated with the null function $f(.)=0$. 
Any targeting function $f$ satisfies two axioms:
\begin{itemize}
  \item%($i$)
    \textbf{monotonicity}: $\comb\subseteq \comb' \implies f(\comb)\leq f(\comb')$. 
  \item%($ii$)
    \textbf{input-sensitivity}: $\exists \comb,\comb' \midwor{s.t.} f(\comb) \neq f(\comb')$.
\end{itemize}
Monotonicity simply reflects that an account with strictly more interest or hobbies should in theory be relevant to more ads, and never to less. Input sensitivity prevents the degenerate case where a targeting function is constant.

A \emph{family} $S$ of \emph{size} $l$ is any collection of $l$ distinct combinations. The \emph{order} of this family is defined as the largest order of a combination it contains. For any family $S$, one can define a targeting function that takes value $f(\comb)=1$ whenever the subset $\comb$ contains at least one combination in $S$.
\ifnum\isTR=1
We now show the converse also holds.
\else
Indeed, as shown in \cite{xray-tr}, the converse is true:
\fi
\begin{lem}\label{res:corefamily}
  For each monotone, input-sensitive targeting function there exists a unique family $S$ satisfying:
  %\begin{itemize}
    %\item

    ($i$) $S$ has size $l$ and order $r$ and it \emph{explains} $f$, which means $f(\comb)=1$ holds if and only if $\exists \comb'\in S, \comb'\subseteq\comb$.

    %\item
    ($ii$) No family of size $l'<l$ explains $f$.

    %\item
    ($iii$) No family of order $r'<r$ explains $f$.
  %\end{itemize}
\end{lem}

\ifnum\isTR=1
Hence, associated with each ad and therefore each targeting function is a unique family of input combinations that is targeted. We call this the ad's \emph{core family}. 

Before proving the result above, we discuss its meaning and consequences.
Let us first introduce a definition, the following order relation will play an important role: We say that a family $S$ \emph{explains} another $S'$ if for any combination $\comb'$ in $S'$ there exists a combination $\comb\in S$ such that $\comb\subseteq\comb'$. Note that according to the definitions above, $S$ explains a function $f$ if and only if it explains $S_{f}= \lset \comb \dimset f(\comb)=1\rset=f^{-1}(\{1\})$ and $S \subseteq S_f$.

For example, with $n=4$ inputs,
$S=\lset\{1,3\}\vf \{4\}\rset$ and
$S'=\lset \{1,2,3\}\vf \{4\}\vf \{2,4\}\vf \{1,3\}\rset$ 
we see that $S$ explains $S'$.
Intuitively, if $S$ explains $S'$, then if we were to observe that all combinations in $S'$ receive an ad, this could in theory be explained by the hypothesis that the ad is targeted at accounts which contain any of the combinations of inputs in $S$. Alternatively, if $S$ does not explain $S'$, then it shows that $S$ is not sufficient on its own to interpret this observation. Similarly, a family $S$ explains $f$ if all its combinations are relevant to the ad, and for any subset of inputs $S'$ that leads $f$ to take value $1$, at least one combination in $S$ is included in $S'$.

Note that, by definition $S_f$ explains $f$, but it does not explain $f$ \emph{succinctly}. In particular $S_f$ is a big family that contains a lot of combinations, and since by monotonicity we have $f(\{1,\ldots,N\})=1$ then $S_f$ contains the combination of all inputs (which has order $N$)). What Lemma~\ref{res:corefamily} and the definition of a core family indicate is that it is possible to find a \emph{small} family, as small as possible both in terms of number and length of combinations involved, that also explains $f$. Note that this result is a consequence of the monotonicity axiom and does not hold for non-monotonic function.

Take the following example: if $f(S)=1$ if and only if $S$ contains a particular input $D_i$. $S_f$ contains all supersets of $\{D_i\}$, a family containing $2^{N-1}$ combinations, but the family $S=\{ \{D_i\} \}$ explains $f$ as well, it is of size 1 and order 1. 

\begin{proof}
Let $\overrightarrow{D}_f$ be the digraph with vertex-set $S_f$ and with arc-set $\lset (\comb,\comb') \dimset \comb \subsetneq \comb' \rset$.
We have that $\overrightarrow{D}_f$ is a DAG because the subset-containment relation defines a partial order.
So, let $S$ be the non-empty set of combinations with null in-degree in $\overrightarrow{D}_f$.
By construction, $S$ explains $S_f$ and $S \subseteq S_f$, hence $S$ explains $f$.

Furthermore, we claim that $S$ is contained in \emph{any} family $S'$ explaining $f$: indeed, since $S'$ is required to contain a subset of any combination $\comb \in S$, and no combination of $S_f$ is strictly contained in $\comb$, then it must contain $\comb$. This shows that $S$ satisfies all conditions of Lemma~\ref{res:corefamily}. Finally, since another family explaining $f$ needs to include $S$, then it will necessarily have a higher size $l$, hence $S$ is the unique with both minimum size and order.
\end{proof}

\iffalse
\subsubsection{Proof of Lemma~\ref{res:corefamily}}
\vspace{-3pt}

The following lemma is used to prove Lemma~\ref{res:corefamily}. First, we say two functions are $r$\_equivalent if they are equal on all combinations of order at most $r$.
\begin{lem}
  The following conditions are equivalent:
  %\begin{itemize}

  (i)~the function $f$ is minimal in its $r$\_equivalent class,

  (ii)~$f^{r}=f$, where $f^{r}$ is defined as

  \hspace{2ex} $f^{(r)}: \comb \mapsto\max\left( f(\comb') \right.\left| \comb'\subseteq \comb\vf |\comb'|\leq r \right)$. 

  (iii)~$S^{(r)}_f$ explains $S_f$, where $S^{(r)}_f$ is defined as

  \hspace{2ex} $S^{(r)}_{f} = \lset \comb \dimset f(\comb)=1 \midwor{and} |\comb|\leq r\rset$,

  (iv)~$f$ is explained by a family of order $r$.
  %\end{itemize}
\end{lem}
\begin{proof}
  Assertion \emph{(i)} implies \emph{(ii)} that implies \emph{(iii)} that implies \emph{(iv)} according to simple properties of $f^{(r)}$: first, observing that by definition $f\geq f^{(r)}$ and second that $f^{(r)}$ is explained by $S^{(r)}_f$.

Assertion \emph{(iv)} implies \emph{(i)} as follows: if $f$ is of order $r$ and $g$ is $r$\_equivalent to $f$, it suffices to show that $f(\comb)=1 \implies g(\comb)=1$ to conclude. Since $\comb\in S_f$, we know there exists $\comb'$ or order at most $r$ such that $f(\comb')=1$ and $\comb'\subseteq\comb$. By monotonicity, $g(\comb)\geq g^{(r)}(\comb) \geq g^{(r)}(\comb') = f^{(r)}(\comb')=f(\comb')=1$, which proves the result.
\end{proof}

%\paragraph{Completing the proof}
Equipped with these definitions and results, we can prove Lemma~\ref{res:corefamily}. 
Given a targeting function $f$, it can be explained by at least one family, namely $S_f$.
Let $l$ be the minimum size over the families explaining $f$.
Similarly, let $r$ be the minimum order over the families explaining $f$.
By the hypothesis we know the function admits one explaining family $S_1$ of minimum size $l$ and one $S_2$ of minimum order $r$. We need to construct an explaining family that satisfies both conditions at the same time. Here is how: Since $S_1\subseteq S_f$ and $S_2$ explains $S_f$, then $S_2$ explains $S_1$. In fact, we can associate with each combination $\comb$ in $S_1$ one chosen in $S_2$ that is contained in $\comb$, and hence we create a new family $S$. We claim it has all these properties.

  First, since $S$ is composed of elements in $S_2$ its order is at most $r$. Second, since $S$ contains one combination for each element in $S_1$ it has at most $l$ combinations. Finally, by construction $S$ explains $S_2$, which by transitivity implies that it explains $S_f$, and hence since it is also contained in it, it explains $f$.

  $S$ satisfies all conditions, for minimum value of $l$ and $r$. It only remains to show that it is unique. 
By contradiction, suppose that another family $S'$ has the same properties.
Since we have that $S,S' \subseteq S_f$, then $S$ explains $S'$ and vice-versa.
By the minimality of $l$, we can furthermore note there is a one-to-one mapping from $S$ to $S'$ satisfying every combination of $S$ explains its image in $S'$ by the mapping.
Since $S$ and $S'$ are not equal, it implies that one element of $S$ is strictly contained into its image in $S'$ and so, we have:
$$\sum_{\combs \in S} |\comb| < \sum_{\comb' \in S'}|\comb'|.$$
But the same argument works in the other direction \textit{i.e.}, from $S'$ to $S'$, that would lead to a contradiction.
\else
Hence, associated with each ad and therefore each targeting function is a unique family of input combinations that are targeted, called the ad's \emph{core family}, and we now sketch why it is correctly identified by our algorithm.

\fi
\fi

\subsection{Algorithm and correctness}
\vspace{-3pt}
\label{s:correctnessproof}

We first describe the gist of the proof of Theorem~\ref{thm:set-intersection} as following from two main claims. These claims are established by using properties of random subsets of elements, which we analyze before providing a formal complete proof.

\subsubsection{Definitions and proof overview}

A subset of inputs $\comb$ is an \emph{$x$\_intersecting subset} of a family $S$ (for $0\leq x\leq 1$) if at least a fraction $x$ of the subsets in $S$ intersect $\comb$ (\ie each contains an input chosen in $\comb$): 
\[
\lset
\combs \in S \dimset \exists i\in \comb, i\in \combs
\rset \geq x\cdot |S|
\ff
\]
Similarly, we say that $S'$ is an \emph{$x$\_intersecting family} of a family $S$ if at least a fraction $x$ of the subsets contained in $S$ contain a combination chosen in $S'$:
\[
\lset
\combs \in S \dimset \exists \comb\in S', \comb \subseteq \combs
\rset \geq x\cdot |S|
\ff
\]
One can immediately deduce the following lemma
\begin{lem}\label{lem:intersect}
Let $S'$ be an $x$\_intersecting family of $|S|$, $\exists \comb$ an $x$\_intersecting subset of $S$ such that $|\comb|\leq|S'|$.
\end{lem}
Indeed, one can build $\comb$ by including for each combination of $S'$ any single input it contains. 

\paragraph{Overview of the proof:}
The gist of the argument for Theorem~\ref{thm:set-intersection} is an original connection between small intersecting subsets and the effect of a core family. Given an ad, let us denote by $\famad$ the family of all inputs combinations that are receiving the ad.  
The proof relies on the following claim: There exists a value of $0<x<1$ such that with high probability an $x$\_intersecting subset of order $\leq l$ exists for $\famad$ if and only if the ad is targeted with a core family of size $\leq l$. Hence, finding such subset is a sound and complete test for detecting that targeting occurs. 

The proof unfolds with two complementary claims: 
\begin{itemize}
	\item \textbf{Completeness:} Let $\famcore$ be the ad's core family, then an $x$\_intersecting subset of $\famad$ with size $|\famcore|$ exists. 
\end{itemize}
This claim holds trivially when targeting is strict and the ad is never shown outside the target (\ie $\pout=0$). Indeed, all combinations of inputs $\combs$ seeing the ad (\ie in $\famad$) necessarily need to be within the target (\ie $f(\combs)=1$) and hence they have to include a combination chosen in $\famcore$. We then deduce that $\famcore$ is a $1$\_intersecting family of $\famad$ with size $l$, hence by Lemma~\ref{lem:intersect} a $1$\_intersecting subset exists with at most the same size.

When targeting is not strict (\ie $\pout>0$) it is more complicated, we can however prove that a similar claim holds for a smaller value of $x$ with high probability by exploiting properties of random subsets as shown below.
\begin{itemize}
	\item \textbf{Soundness:} If targeting does not occur, then $\famad$ does not admit an $x$\_intersecting subset of size $l$.
\end{itemize}
	This claim follows naturally from the properties of random subsets and is not qualitatively surprising. However, it is important that we prove that the same value of $x$ that is used for completeness also allows to obtain that property, which is why a careful analysis is required. Note also that it is critical that both properties hold using a small number $m$ of accounts to test (\ie $m$ should be of the order $\ln(N)$ where $N$ is the number of inputs to monitor).

% The key property to explain our algorithm is random subsets. We can show under the conditions of the theorem that there exists $0<x<1$ that satisfies two properties related to the inputs of accounts receiving the ads: (1) if targeting does not occur, then with a large probability we cannot find a subset of $l$ inputs that meets at least a fraction $x$ of the accounts seeing the ad, and (2) if targeting does occur, we have accounts receiving the ads for various reasons, within and outside the targeting scope; but we can show with high probability that at least a fraction $x$ of them are within scope and hence must include one combination in the core family. Since with each core family of size $l$ one can associate an intersecting subset that contains at most $l$ elements, checking the existence of such a subset reveals the presence of targeting.

\paragraph{}
Finally, while the above argument explains an algorithm can \emph{detect} that targeting takes place, it does not explain how the core family can be \emph{exactly computed}. Again, this can be done by leveraging stronger results of random subsets, and we present different algorithms that determine the core family with varying time-complexity tradeoff. 

\subsubsection{Random subsets and probabilistic inequalities}

Let us start with some definitions:
\begin{itemize}
   \item 
    A random \emph{Bernoulli subset}, denoted by $B(n,p)$, is a subset such that any of $n$ elements is contained with probability $p$ independently of all others.
 \item 
  A random \emph{Bernoulli family} of size $m$ is a collection of $m$ independent Bernouilli subsets.
\end{itemize}

Since Bernouilli subsets and families derive from many independent decisions to include or not a single element, we will use inequalities on the distribution of sum of binary variables, especially this one due to Chernoff:
\begin{lem}\label{res:chernoff}
	If $Y$ is a sum of independent binary variables, let $\mu=\expec{Y}$, we have for any $0<\delta\leq 1$:
	\[
	\begin{array}[c]{l}
		\probaof{Y\geq (1+\delta)\mu} \leq \exp\left( -\frac{\delta^2 \mu}{3} \right)\midwor{, and} \\	
		\probaof{Y\leq (1-\delta)\mu} \leq \exp\left( -\frac{\delta^2 \mu}{2} \right) \\	
	\end{array}
	\]
\end{lem}
Thus, for any polynomial $P$, integer $N$ and value $\varepsilon>0$, 
% this implies that with probability $\left( 1-{\varepsilon}/{P(N)} \right)$ $S$ is within distance $\delta$ of its mean if we have:
\[
\mu \geq \frac{3}{\delta^2} \ln\left( \frac{2 P(N)}{\varepsilon} \right)
\implies
\probaof{|Y-\mu|\leq \delta\mu} \geq 1-\frac{\varepsilon}{P(N)}
\ff
\]
In other words, such variable $Y$ remains close to its expectation (\ie up to a constant multiplicative factor) except on an event of polynomially small probability. This holds as soon as its expectation is at least logarithmic. 

The lemma below allows us to prove soundness:
\begin{lem}\label{res:soundness}
  Let $1>x>0$, $l\in\NatInt$, 
  \(
  p < 1-(1-x)^{\frac{1}{l}}\;,
  \) 
  and a Bernouilli family $B_1(n,p),B_2(n,p),\ldots,B_m(n,p)$.
  There exists $C>0$ such that for any $\varepsilon>0$ and polynomial $P$, if 
\ifnum\isTR=0\vspace{-8pt}\fi
  \(
  m\geq C \cdot
  %\frac
  {\left( l\ln(n)+\ln P(n)+\ln(1/\varepsilon) \right)}
  %{\ln\left( {x^x (1-x)^{1-x}}/{\left( 1-(1-p)^l \right)^x} \right)}
  \vf
  \)
\ifnum\isTR=0\vspace{-8pt}\fi
  then with probability $\left( 1-{\varepsilon}/{P(n)} \right)$ no $x$\_intersecting subset exists of size $l$ for this Bernouilli family.
\end{lem}

\ifnum\isTR=1
\begin{proof}
Let us consider an arbitrary subset $\comb$ of size $l$.
The probability that it intersects an arbitrary Bernouilli subset is $1 - (1-p)^l$. If we introduce $Y$ the variable counting how many Bernouilli subset $\comb$ intersects, we observe that it is a sum of binary independent variables, with expectation $\mu= \left( 1 - (1-p)^l \right)m$. We also note that $\comb$ is an $x$\_intersecting subset exactly if $Y\geq xm$. Assuming $p < 1-(1-x)^{\frac{1}{l}}$ as we do, $\mu$ is multiplicatively smaller than $xm$. Hence we can apply Chernoff Bound to conclude that
$\probaof{Y\geq xm}\leq\frac{\varepsilon}{P(n) n^l}$ when
%$\comb$ is an $x$\_intersecting subset with probability at most $\frac{\varepsilon}{P(n) n^l}$ if
\[
  m\geq C \cdot
  \left( \ln\left(n^l P(n) /\varepsilon \right) \right)
  ,\midwor{
    with  $C=3\frac{1-(1-p)^l}{\left(x-\left( 1-\left( 1-p \right)^l \right)\right)^2}$.
  }
\]
%The constant $C$ depends on the difference between $x$ and $1 - (1-p)^l$.
Since there are ${\binom n l}\leq n^l$ choices of $\comb$, by the union bound the probability that at least one of them is an $x$\_intersecting subset is at most $\frac{\varepsilon}{P(n)}$.
\end{proof}

\fi
\ifnum\isTR=1
\subsubsection{Detailed proof}

First, let us consider soundness. Assuming no targeting takes place, subsets of inputs in $\famad$ are chosen independently of the inputs that they contain. Hence it is a Bernouilly family of average size $p_{\emptyset}m$ with parameter $N$ (the number of inputs) and $p=\alpha$. By choosing $x>1-(1-\alpha)^l$ and $m$ sufficiently large, with very high probability no $x$\_intersecting subset of size $l$ exists. In this case, our test correctly concludes that no targeting is taking place.

Now, let us consider completeness. We already explained why this test will be correct when the ad is received only by accounts within the target (\ie $\pout=0$) but it remains to be shown in the general case. We start from the following observation: The family $\famad$ is composed of two families. The first, $\famadin$, contains subsets of inputs that are in the target, and hence contain a combination of $\famcore$. The second, $\famadout$, includes subsets that are not in the target but received the ads due to $\pout>0$. It can be observed that the size of both families depends on the values of $\pout$, $\pin$, $\alpha$. We already know that a $1$\_intersecting subset of size $l$ exists for $\famadin$, that we can construct using $\famcore$. Note that it is also an $x$\_intersecting subset for $\famad$, where $x = |\famadin|/(|\famadin|+|\famadout|)$. 

\begin{lem}\label{res:superset}
	We assume that targeting takes place, where the targeting function admits a core family of size $l$ and order $r$, and uses targeting probability $\pin$ and $\pout$. Let $x>0$, $\alpha>0$, and assume
  \(
  \pout/\pin < \frac{1-x}{x} \frac{\alpha^r}{1-\alpha^r}
  \;.
  \)
Finally, let $\comb$ be any combination. 
  
  There exists $C>0$ such that for any $\varepsilon>0$ and polynomial $P$, whenever we have
  \[
  m\geq C \cdot
  %\frac
  {\left( \ln P(N)+\ln(1/\varepsilon) \right)}
  %{\ln\left( {x^x (1-x)^{1-x}}/{\left( 1-(1-p)^l \right)^x} \right)}
  \;.
  \]
  then with probability $\left( 1-{\varepsilon}/{P(N)} \right)$ the following holds: among accounts containing $\comb$ and receiving the ad, at least a fraction $x$ of them is within the targeting scope, 
  \[
  \ie
  \frac{
  \left|
  \lset\combs\in\famadin \dimset \comb\subseteq \combs\rset
  \right|
  }{
  \left|
  \lset\combs\in\famad \dimset \comb\subseteq \combs\rset
  \right|
  }
  \geq x 
  \ff
  \]
\end{lem}
\else
To prove property (2), we need to bound, among accounts receiving an ad, the fraction that is outside the scope of targeting but still receives the ads because $\pout>0$. Formally, we have:
\begin{lem}
  Let $x>0$, $\alpha>0$, and a core family of size $l$ and order $r$. Let $\pin$, $\pout$ be such that
  \(
  \pout/\pin < \frac{1-x}{x} \frac{\alpha^r}{1-\alpha^r}
  \;,
  \)
   and let $\comb$ be a combination of order (at most) $r$. 
  
  For any $\varepsilon>0$ and polynomial $P$ of degree $\leq r$, there exists $C>0$ such that with probability $\left( 1-{\varepsilon}/{P(n)} \right)$ the following holds: among accounts containing $\comb$ and receiving the ad, at least a fraction $x$ of them is within the targeting scope whenever we have:
\vspace{-8pt}
  \[
  m\geq C \cdot
  %\frac
  {\left( r\ln(n)+\ln(1/\varepsilon) \right)}
  %{\ln\left( {x^x (1-x)^{1-x}}/{\left( 1-(1-p)^l \right)^x} \right)}
  \;.
  \]
\end{lem}
\vspace{-8pt}
\fi

\ifnum\isTR=1
\begin{proof}
For each of the accounts $A_1,\ldots,A_m$ we introduce $Y_j$ which takes he following value:
\[
\left\{
\begin{array}[l]{ll}
	1 &\midwor{if $A_j$ is in target, sees the ad, and $\comb\subseteq A_j$,} \\
	-\frac{x}{1-x} &\midwor{if $A_j$ is not in target, sees the ad, and $\comb\subseteq A_j$,}\\
	0 &\midwor{otherwise} \\
\end{array}
\right.
\]
We then introduce $Y = \sum_{j=1}^m Y_j$, which is a sum of binary independent variables. We also note that the property of the theorem holds exactly if $Y\geq 0$, it is then sufficient to prove that this occurs with high probability using a Chernoff bound argument.

First by the linearity of expectation we have that: 
$$
\begin{disarray}[l]{ll}
\mathbb{E}[Y] & = \sum_{j=1}^m \left( \alpha^{|\comb|} q_{\comb} p_{in} - \frac x {1-x} \alpha^{|\comb|} \left(1 - q_{\comb}\right) p_{out} \right)
\\ & = \left(q_{\comb} p_{in} - \frac x {1-x} \left(1 - q_{\comb}\right) p_{out}\right)\alpha^{|\comb|}m,
\end{disarray}
$$ 
where $q_{\comb}$ denotes the probability for an account to be within scope knowing that it contains $\comb$.
This expectation is positive as long as it holds that
  \(
  \pout/\pin < \frac{1-x}{x} \frac {q_{\comb}} {1-q_{\comb}}
  \;.
  \)
Moreover, the above upper-bound is monotonically increasing with $q_{\comb}$, which is at least $\alpha^r$ because it suffices to complete $\comb$ with any combination of the core to be within scope.
As a result, it always holds that $\expec{Y} > 0$ (with respect to our assumption about the ratio $\pout/\pin$ for the lemma).

Accordingly we deduce 
whenever $m\geq C \cdot \ln\left( P(N)/\varepsilon\right)$ 
  \[
%   m\geq C \cdot
%   %\frac
%   { \ln\left(n^r/\varepsilon\right) }
%   %{\ln\left( {x^x (1-x)^{1-x}}/{\left( 1-(1-p)^l \right)^x} \right)}
%   ,\midwor{
%\midwor{with}
\begin{disarray}[c]{rl}
	\midwor{with}
  C= &
  \frac{2}{\alpha^{|\comb|} q_{\comb} \pin}
  \left({1-\frac{x}{1-x}\frac{1-q_{\comb}}{q_{\comb}} \frac{\pout}{\pin}} \right)^{-1} \\
  \leq & 
  \frac{2}{\alpha^{|\comb|+r}\pin}
  \left({1-\frac{x}{1-x}\frac{1-\alpha^r}{\alpha^r} \frac{\pout}{\pin}} \right)^{-1}\vf
\end{disarray}
  \]
that $\probaof{Y\geq 0}\geq 1-\frac{\varepsilon}{P(N)}$ holds 
% then the random variable $Y$ is non-negative with probability at least $\left( 1-{\varepsilon}/{P(n)} \right)$.
% The constant $C$ depends on $\left(q_{\comb} p_{in} - \frac x {1-x} \left(1 - q_{\comb}\right) p_{out}\right)$.  
\end{proof}

\fi
%Property (2) is a corollary of this lemma with $\comb=\emptyset$. 

\ifnum\isTR=1
\paragraph{Final argument.}
According to Lemma~\ref{res:superset} (applied with $\comb=\emptyset$), the existence of an $x$\_intersecting set of size $l$ is guaranteed with high probability, if we can satisfy the condition on $x$. 

In particular, since we could a priori fix $\alpha$ to satisfy $x>1-(1-\alpha)^l$ we have that both proof apply simultaneously whenever there exists $0<x<1$ verifying:
\begin{equation}\label{eq:tagratio}
	\frac{\pout}{\pin} < \frac{1-x}{x} \frac{(1-(1-x)^{\frac{1}{l}})^r}{1-
	(1-(1-x)^{\frac{1}{l}})^r} = \varphi_{l,r}(x) \ff
\end{equation}
Whenever this condition is verified (\ie whenever the gap between $\pout$ and $\pin$ is sufficiently large), one can choose a value of $x$, $\alpha$ and subsequently $C$ such that if $m\geq C \ln(n/\varepsilon)$ the detection test is correct with probability $1-\varepsilon/n$.

Note that while finding an $x$\_intersecting subset is a sufficient evidence that targeting takes place, it does not allow us to directly compute the core family. In particular this subset is neither a combination of the core family, it is a union of elements that all appear in at least one combination of the core family, but it is not unique.

However, using this detection brick, various algorithms can be used to exhaustively search for a core family. We will also show that a polynomial-time algorithm can refine this analysis to compute the core family at the expense of a more complex recursion.

\subsubsection{When is the condition verified?}

The condition of Eq.(\ref{eq:tagratio}) is important because it denotes the maximum ratio $\pout/\pin$ that can be detected by our algorithm. Intuitively, if this ratio is $1$ and $\pout=\pin$ targeting has no effect and hence its presence and its core family remains impossible to determine. Since the choice of the percentage $x$ is a parameter of the algorithm that can be tuned (along with the value of $\alpha$) it would be interesting to know under which condition we can detect targeting with the largest $\pout/\pin$ ratio. The following lemma answers that question precisely:

\begin{lem}
	Let $M_{l,r}=\max_{x\in ]0;1[ } \varphi_{l,r}(x)$, we have
	\[
	\left\{
	\begin{array}[c]{ll}
		\midwor{if} 
		l=1, & M_{1,r} %= \lim_{x\to 1^-} \varphi_{1,r}(x) 
= {1}/{r}\vf \\
		\midwor{if} 
r=1, & M_{l,1} %= \lim_{x\to 0^+} \varphi_{l,1}(x)
= {1}/{l}\vf \\
		\midwor{if} 
r=l=n>1, & M_{n,n} = 1/(2^n-1)^2 \vf \\
		\midwor{if} 
r>1,l>1, & M_{l,r} = \frac{z^l}{1-z^l}\frac{(1-z)^r}{1-(1-z)^r}\vf
	\end{array}
	\right.
	\]
where $z$ is the only solution in ]0;1[ of
\[
r z^{l+1}-{l}(1-z)^{r+1} - (r+{l})z+{l} =0
\vf
\]
and this maximum is attained for $x=1-z^l$.
\end{lem}

\begin{proof}
	When $l=1$ one can easily see that $\varphi_{1,r}$ is strictly increasing on this interval and computes its limit as $x$ approaches $1$. A similar argument holds for $r=1$. 

	Whenever $r>1$ and $l>1$, introducing the new variable $z=(1-x)^{1/l}$ we first observe:
	\[
%\max_{x\in [0;1] } 
\varphi_{l,r}(z)
= 
%\max_{z\in [0;1] } 
f_l(z) \cdot f_r(1-z)
\vf
\textrm{where $f_n(z)=\frac{z^n}{1-z^n}$.}
\]

We observe 
$\varphi_{l,r}'(z)=
f'_l(z) \cdot f_r(1-z)
- f_l(z) \cdot f'_r(1-z)
$, and note that this derivative becomes null whenever $f'_l(z)/f_l(z)=f'_r(1-z)/f_r(1-z)$. Moreover, we have
\[
f'_n(z)=-\frac{l z^{n-1}}{(1-z^{n})^{2}}
\midwor{hence}f'_n(z)/f_n(z)=\frac{n}{z(1-z^n)}
\]
so that the condition is $\frac{l}{z(1-z^l)} = \frac{r}{(1-z)(1-(1-z)^r)}$ which yields the value of $z$ reaching the maximum.

To conclude, we just need to observe that there is a unique solution in $]0;1[$. We can immediately observe, when $r>1$ and $l>1$ that the product $f_l(z) \cdot f_r(1-z)$ has null limits on both side, and a derivative that is positive near $0^+$ and $1^-$. Since its third derivative is strictly positive, its second derivative increases and can only be null once. We deduce that the derivative cannot cancel twice between $0$ and $1$ since it would create two inflexion points.

Finally, when $r=l=n$, since the product is symmetric in $z$ and it has a unique maximum on $]0;1[$ it has to be in $z=\frac{1}{2}$ which yields the result.
\end{proof}

According to this lemma, when a single input is used for targeting \textit{i.e.}, $l=1,r=1$, the condition is always verified as soon as $\pout<\pin$ and hence any targeting is detected. When the targeting uses a single combination of $r>1$ inputs (\ie $l=1$) or a union of $l>1$ single inputs (\ie $r=1$), the condition holds as long as $\pout$ is below some threshold. 

When $l$ and $r$ are allowed to grow beyond $1$, the quick combinatorial explosion of the number of hypotheses to test by our system requires that the ratio $\pout/\pin$ decreases exponentially fast, but detection remains possible. For $l=r=3$, a relatively complex case, we can still detect targeting even when $2\%$ of accounts outside the target received the ads.
\begin{figure}[h]
	\begin{center}
		\includegraphics[width=0.99\linewidth]{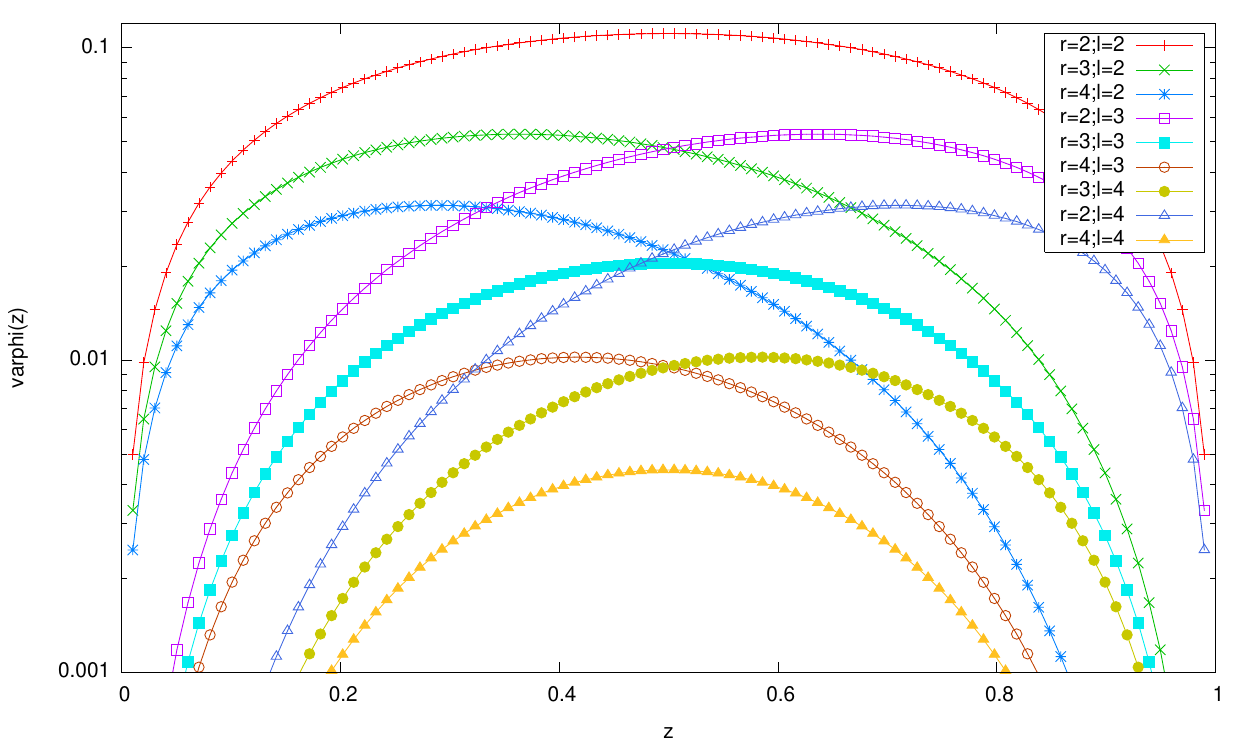}
	\end{center}
	\caption{The function $\varphi_{l,r}(z)$ for $l>1$ and $r>1$.}
	\label{fig:varphi}
\end{figure}
Figure~\ref{fig:varphi} presents the value of the RHS defining the necessary condition, as a function of the variable $z=(1-x)^{\frac{1}{l}}$. We observe maxima for different values of $l$ and $r$.

\subsubsection{Beyond detection, computing the core family}

So far, we have shown that, after computing the family made with inputs of account receiving ads, looking for an $x$\_intersecting subset of this family with size $l$ is a correct test with high probability whenever we have a logarithmic number of accounts. If this test determines that targeting does not take place, there is no other explanation to find. However, if targeting occurs, one would also like to deduce from this test \emph{which} combination of inputs are used for targeting this ad, or computing exactly the core family of the function $f$.

Here we show that under the same condition as detection, we can compute the core family. There are multiple algorithms to do so, each one potentially better depending on what is known about the targeting. They all use a common result that we draw below:

\begin{lem}
	We assume Eq.(\ref{eq:tagratio}) and targeting occurs with a core family $\famcore$. There is $C>0$, $0<x<1$ such that for any $\varepsilon>0$, polynomial $P$, and combination $\comb$, if 
	$m\geq C\cdot \left( \ln(n) + \ln P(n) + \ln(1/\varepsilon)\right)$, then with probability $(1-\frac{\varepsilon}{P(n)})$ exactly \underline{one} of the following claims holds:
% Let $\comb$ be a combination of inputs, and 
% 	\[
% 	\lset \combs \in \famad \dimset \comb\subseteq\combs \rset
% 	\midwor{and}\]
	\begin{itemize}
		\item[]$(i)$ $\comb$ contains a combination from the core family 
			\[\ie \exists \combs\in\famcore\vf \combs\subseteq\comb\ff\]
		\item[]$(ii)$ an $x$\_intersecting subset of size $l$ exists for %$\famdif{\comb}$,
	\[
	%\textrm{where}\;
	\famdif{\comb} =
	\lset \combs\cap\overline{\comb} \dimset \combs \in \famad\vf \comb\subseteq\combs \rset
	%\midwor{then}
	\ff\]
	\end{itemize}
\end{lem}
This result combines all lemmas used in the proof of the detection test. In fact, with the convention $\famcore=\emptyset$ used to denote non-targeting, it contains the proof of detection test as a particular case with $\comb=\emptyset$. But its strength is to be applied to multiple different combination $\comb$ as a building block to determine $\famcore$.

\begin{proof}
	First we prove $(i)$ implies $(ii)$ cannot hold which is the easy part of the result. If a combination of the core is contained in $\comb$, then any account that contains $\comb$ as part of its input is in the target and hence it receives an ad with probability $\pin$, and this holds irrespectively of all other inputs. One deduces that $\famdif{\comb}$ in that case is a Bernouilli family, we can then apply Lemma~\ref{res:soundness} and conclude that $(ii)$ may only occur with probability ${\varepsilon}/{P(n)}$.

	We now show that if $(i)$ does not hold, then $(ii)$ does. %We first introduce
	\[
	\textrm{Let}\;
	\famdifin{\comb} =
	\lset \combs\cap\overline{\comb} \dimset \combs \in \famadin\vf \comb\subseteq\combs \rset\vf
	\]
	\[
	\textrm{and}\;
	\famdifcore{\comb} =
	\lset \combs\cap\overline{\comb} \dimset \combs \in \famcore \rset\ff
	\] 
	Note that since no combination of the core family is included in $\comb$, no element of $\famdifcore{\comb}$ is empty. Observe that, by definition a combination in $\famadin$ should contain a combination of the core. This directly implies that a combination in $\famdifin{\comb}$ necessary contains a combination from $\famdifcore{\comb}$, which is by consequence a $1$\_intersecting family of $\famdifin{\comb}$.
	
	It is an immediate consequence of Lemma~\ref{res:superset} that under the condition above, $|\famdifin{\comb}|/|\famdif{\comb}|\geq x$ holds with probability $1-\varepsilon/P(n)$. 
	Therefore $\famdifcore{\comb}$ is an $x$\_intersecting family of $\famdif{\comb}$, proving $(ii)$.
\end{proof}

\paragraph{}
The result above shows that under the same conditions as those used for detection, one can design a provably correct test to decide whether a combination $\comb$ is a superset of a combination in the core. This test resembles the previous one, it looks for an intersecting subset of size $l$ that does not use the inputs of $\comb$ among the accounts containing $\comb$. It uses no more than $O(N^{l+1})$ operations with a naive exhaustive search. What remains to be shown is how one can conduct multiple tests on various combinations $\comb$ to compute $\famcore$. There are multiple ways:

\paragraph{Agglomerative algorithms:}
% If $\comb$ is part of the core family, then removing $\comb$ from the subsets in $S^{(ad)}_{\comb}$ yields a random Bernouilli family whose average size depends on $\alpha,r,p_{in}$ and $m$.
% So, we deduce from Lemma~\ref{res:soundness} that assuming $m$ is chosen large enough, every $x\_$intersecting subset for this family has to intersect $\comb$, almost asymptotically surely.
% The same also holds if $\comb$ is a superset of some combination into the ad's core family.
% 
% Conversely, if $\comb$ does not contain any combination from the core, then in particular, every combination of the ad's core family contains an input which is not $\comb$.
% By Lemma~\ref{res:superset}, if we choose $m$ large enough then the union of (at most) $l$ such inputs, namely one for each combination of the core family, yields w.h.p. another $x\_$intersecting subset for the accounts in $S^{(ad)}_{\comb}$ that does not intersect $\comb$.
% 
% To sum up, w.h.p. we have that $\comb$ is a superset of some combination in the ad's core family iff it intersects every $x\_$intersecting subset for the accounts in $S^{(ad)}_{\comb}$.
%This property can be checked in $O(N^{l+1})$ time for every of the $O(N^{r'})$ combinations we consider and so, it takes $O(N^{r'+l+1})$ time to compute all of the supersets of the combinations in $S^{(core)}$.
%The computation of $S^{(core)}$ follows from the selection of all the inclusionwise minimal combinations amongst these supersets.
Assume an upper bound $\lmax$ is known. A simple (costly) search looks for the results of all tests for all combinations. For instance, one can maintain a current core $\famcore$ initialized to be empty, and a queue of combinations remaining to be checked, which is initialized to contain $\comb=\emptyset$. The first test in effect tests whether targeting occurs. Whenever a combination $\comb$ is at the head of the queue, we update it as follows: (1) if the combination already contains one combination identified in $\famcore$, simply drop it otherwise run the test; (2) if the test finds an intersecting subset of size $l\leq \lmax$, conclude that $\comb$ does not contain a combination of the core, and add $N-|\comb|$ combinations to the queue constructed as $\lset \comb\cup\{i\} \dimset i\notin\comb \rset$, while avoiding those already in the queue; (3) if the test concludes that $\comb$ contains a combination of the core, add $\comb$ to $\famcore$.

It's possible to run the queue infinitely, stopping whenever $\lmax$ combinations have been identified, or when all combinations of order $\rmax$ have been checked, assuming such bound is known. This uses at most $O(N^{\lmax+\rmax+1})$ operations.

\paragraph{Removal algorithms:}
There are two drawbacks in the precedent algorithm: it tests a large number of combinations, and if the bound $\rmax$ is loose, and $l<\lmax$ it will test absolutely \emph{all} combinations of size $\rmax$ before concluding, which seems very costly. We now present another algorithm that does not assume any bound on $r$, and prevents this exhaustive search. 

It works as follows: let us assume we already identified a family of some combinations in the core, $S\subseteq \famcore$. If we assume we start from a combination $\comb$ that (1) is not a superset of a combination in $S$, and (2) contains a combination from the core family $\famcore$, then we are guaranteed to find another combination of $\famcore$ using at most $|\comb|$ tests. In fact, one can update $\comb$ as follows: order all inputs from $\comb$ arbitrarily, and for each one do the following: first remove the input from $\comb$ and run the test to determine whether it still contains a combination of the core. If the test indicates that a core combination remains, this removal is permanent, otherwise, it proves that, for the remainder of inputs left, this one is ``critical'' and we put it back in $\comb$. After we do that for all inputs, the ones remaining in $\comb$ form a combination of the core.

One can start with $S=\emptyset$ and $\comb$ containing all inputs, as this is guaranteed to find a core combination $\comb_1$. At any time, $S$ contains at most $l$ combinations of at most $r$ inputs, which means there are at most $r^l$ subsets constructed by taking all inputs and removing at least one inputs from each of the combinations of $S$. All those subsets satisfy property (1) above, but not necessarily (2). In fact, we can consider them in any order, and run the test of property (2). If one of them does satisfy it, it can be used to find a new combination of the core, and the process repeats with a new value of $S$. Otherwise, if all of those subsets are shown not to contain any more combination, we can conclude that $S$ contains all combinations of the core. There could not be more than $l$ combinations in $S$, hence this algorithms uses at most $l r^l N$ tests, which hence uses in total $O(N^{\lmax+2})$ operations.

\else
The two lemmas above (proved in \cite{xray-tr}) can be combined whenever $\alpha$ satisfies the inequality for $p$ in the first lemma, which shows that an algorithm can detect the presence of targeting whenever  
\(\pout/\pin < \frac{1-x}{x} \frac{(1-(1-x)^{\frac{1}{l}})^r}{1-
(1-(1-x)^{\frac{1}{l}})^r}\).

A naive exponential algorithm could be used to exhaustively search for a core family using this brick. We also show that a polynomial algorithm can refine this analysis to compute the core family at the expense of a more complex recursion in \cite{xray-tr}.
\fi

\end{document}